\documentclass[journal]{IEEEtran} % \documentclass[lettersize,journal]{IEEEtran}
\usepackage{amsmath,amsfonts}
\usepackage{algorithmic}
\usepackage{algorithm}
\usepackage{array}
\usepackage{booktabs}
\usepackage[caption=false,font=normalsize,labelfont=sf,textfont=sf]{subfig}
\usepackage{textcomp}
\usepackage{stfloats}
\usepackage{url}
\usepackage{verbatim}
\usepackage{graphicx}
\usepackage{cite}
\usepackage{standalone}
\usepackage{tablefootnote}
\usepackage{multirow}
\usepackage{siunitx}
\usepackage{xcolor}
\usepackage{tikz}
\usetikzlibrary{
    positioning,
    arrows.meta,
    shapes.geometric,
    calc
}
\usepackage{tikz}
\usepackage{graphicx}
\graphicspath{{figures/images/}{../images/}}
\usetikzlibrary{positioning, arrows.meta, shapes.geometric, calc, backgrounds, patterns, patterns.meta}

\def\imgsize{3.8cm}         % Width/Height of squared images
\def\blockheight{4.5cm}     % Visual width of rotated blocks
\def\mainwidth{1.2cm}       % Thickness of Blue blocks
\def\opwidth{0.8cm}         % Thickness of Orange/Green/Gray blocks
\def\rowgap{3.0}            % Y-coordinate for top/bottom rows
\definecolor{encoder_blue}{RGB}{200, 200, 255}
\definecolor{quant_green}{RGB}{200, 255, 200}
\definecolor{entropy_orange}{RGB}{255, 210, 180}
\definecolor{op_dark_gray}{RGB}{210, 210, 210}
\definecolor{bit_gray}{RGB}{240, 240, 240}
\definecolor{bg_blue}{RGB}{215, 225, 255}
\definecolor{bg_pink}{RGB}{255, 220, 225}

\definecolor{conv_orange}       {RGB}{240, 160, 100}  % Conv (no stride)  — darker
\definecolor{conv_stride_orange}{RGB}{255, 200, 150}   % Conv (stride=2) — warm  orange
\definecolor{act_violet}        {RGB}{220, 195, 255}   % Activations (ReLU) — soft violet
\definecolor{act_gdn_stripe}    {RGB}{140,  90, 190}   % Stripe color for GDN/ReLU blocks
\definecolor{res_blue}          {RGB}{180, 215, 255}   % ResBlocks    — steel blue

\newcommand{\rottext}[2]{%
    \rotatebox{#1}{%
        \begin{tabular}{@{}c@{}}
            #2
        \end{tabular}%
    }%
}

\def\isoCos{0.866}   % cos(30°)
\def\isoSin{0.500}   % sin(30°)

\newcommand{\drawcuboid}[6]{%
    \pgfmathsetmacro{\cbxr}{#1+#3}%          x of front right edge
    \pgfmathsetmacro{\cbyt}{#2+#4}%          y of front top edge
    \pgfmathsetmacro{\cbdx}{#5*\isoCos}%     isometric x-offset
    \pgfmathsetmacro{\cbdy}{#5*\isoSin}%     isometric y-offset
    \filldraw[fill=#6!45, draw=black!30, line width=0.5pt]
        (\cbxr, #2)
        -- ({\cbxr+\cbdx}, {#2+\cbdy})
        -- ({\cbxr+\cbdx}, {\cbyt+\cbdy})
        -- (\cbxr, \cbyt) -- cycle;
    \filldraw[fill=#6!65, draw=black!30, line width=0.5pt]
        (#1, \cbyt) -- (\cbxr, \cbyt)
        -- ({\cbxr+\cbdx}, {\cbyt+\cbdy})
        -- ({#1+\cbdx}, {\cbyt+\cbdy}) -- cycle;
    \filldraw[fill=#6, draw=black!30, line width=0.5pt]
        (#1, #2) -- (\cbxr, #2) -- (\cbxr, \cbyt) -- (#1, \cbyt) -- cycle;
}

\newcommand{\drawcuboidgdn}[5]{%
    \drawcuboid{#1}{#2}{#3}{#4}{#5}{act_violet}%
    \pgfmathsetmacro{\gdnxr}{#1+#3}%
    \pgfmathsetmacro{\gdnyt}{#2+#4}%
    \pgfmathsetmacro{\gdnspan}{#3+#4}%   diagonal coverage span
    \begin{scope}
        \clip (#1, #2) rectangle (\gdnxr, \gdnyt);
        \foreach \gs in {0,0.16,...,\gdnspan}{%
            \draw[act_gdn_stripe, line width=1.6pt]
                (#1 - #4 + \gs, #2) -- (#1 - #4 + \gs + \gdnspan, #2 + \gdnspan);
        }%
    \end{scope}
    \draw[black!30, line width=0.5pt]
        (#1, #2) -- (\gdnxr, #2) -- (\gdnxr, \gdnyt) -- (#1, \gdnyt) -- cycle;
}

\usepackage[hidelinks,breaklinks=true]{hyperref}
\begin{document}

\title{Hardware-Aware Deployment of Joint SAR Compression and Despeckling on FPGA} %FPGA-accelerated SAR Compression and Despeckling}
% original: Towards Onboard SAR Compression: Enhancing SAR Data Compression and Despeckling for FPGA-based acceleration
% Alternative titles discussed with Paolo at EUSAR26: FPGA-based Joint SAR Compression and Despeckling, FPGA-accelerated SAR Compression and Despeckling

\author{Cédric~Léonard, %~\IEEEmembership{Member,~IEEE,} %\orcidlink{0009-0004-2252-2173}
        Francescopaolo~Sica, %~\IEEEmembership{Senior Member,~IEEE,} %\orcidlink{0000-0003-1593-1492}
        and~Martin~Schulz. %~\IEEEmembership{Life~Fellow,~IEEE} % <-this % stops a space %\orcidlink{0000-0001-9013-435X}
\thanks{C. Léonard is with the Technical University of Munich, Munich, Germany and the Remote Sensing Technology Institute, German Aerospace Center (DLR), Weßling, Germany (e-mail: cedric.leonard@dlr.de).} % <-this % stops a space
\thanks{F. Sica is with the Department of Aerospace Engineering, University of the Bundeswehr Munich, Neubiberg, Germany (e-mail: francescopaolo.sica@unibw.de).} % <-this % stops a space
\thanks{M. Schulz is with the Technical University of Munich, Munich, Germany (e-mail: martin.w.j.schulz@tum.de).} %
% TODO(arXiv preprint): comment the \thanks{} line out
% \thanks{Manuscript received April XX, XXXX; revised September XX, XXXX.}
}

% TODO(arXiv preprint): empty the first argument of the \markboth{}{} command
\markboth{} %
{Léonard \MakeLowercase{\textit{et al.}}: FPGA-accelerated SAR Compression and Despeckling}

%\IEEEpubid{0000--0000/00\$00.00\copyright~2021 IEEE} % @TODO: To uncomment and fix, it overlaps with the text.

\maketitle

\begin{abstract}
Next-generation Synthetic Aperture Radar (SAR) missions will generate data far faster than they can downlink, making onboard data reduction essential for near-real-time Earth observation.
Learned Image Compression (LIC) offers better rate-distortion performance than handcrafted codecs used operationally today, and recent work shows that simultaneously despeckling and compressing SAR imagery enables better representation capacity while unlocking higher compression rates. %removes a noise source that otherwise wastes precious representation capacity.
These methods, however, have yet to be confronted with the strict power, compute, and operational constraints of spaceborne systems.
In this work, we bridge this gap by deploying a joint SAR Despeckling and Data Compression (DDC) framework on an embedded ZCU102 FPGA-based platform, introducing model adaptations that respect the accelerator's fixed-point arithmetic and limited set of supported operations. %through the Vitis AI overlay accelerator.
We evaluate four model topologies across precision levels and across CPU, GPU, and FPGA platforms, revealing several findings with direct design implications.
We find that replacing conventional GDN activation functions with plain ReLU improves quality on SAR, suggesting that design principles established for compression of natural images do not necessarily transfer to SAR imagery.
In addition, we demonstrate that residual blocks offer little representational benefit for ten times the compute, and show that the FPGA is the most energy-efficient of the platforms tested.
% We evaluate four model designs across precision levels and across CPU, GPU, and FPGA platforms, revealing several findings with direct design implications: replacing the conventional GDN layers with plain ReLU unexpectedly improves quality on SAR; residual blocks offer little representational benefit for ten times the compute; and the FPGA is the most energy-efficient of the platforms tested.
Together, these results set a functioning edge deployment workflow and an evidence-based starting point for onboard SAR compression.
The code is available at \url{https://github.com/CedricLeon/SAR_DDC_FPGA}.
\end{abstract}


% \fi

\IEEEpeerreviewmaketitle

\section{Introduction}\label{sec:intro}

\IEEEPARstart{T}{he} growing volume of Earth observation data generated by modern satellite missions poses significant challenges for onboard storage and downlink transmission.
Synthetic Aperture Radar (SAR) systems, in particular, produce large volumes of high-resolution data, placing increasing pressure on satellite communication bandwidth.
This motivates the need for efficient onboard data compression.

Beyond the data volume itself, the SAR processing pipeline also influences the amount of information that must be transmitted to the ground segment.
In conventional missions, raw radar echoes are downlinked and processed on the ground to produce focused SAR images.
However, the growing demand for near-real-time applications, such as disaster monitoring, maritime surveillance, and environmental monitoring, requires to migrate parts of the SAR processing chain directly onboard the satellite.
This paradigm is especially relevant for modern missions with limited communication bandwidth, short ground contact windows, or large satellite constellations generating vast amounts of data.

Traditionally, SAR missions rely on specialized compression techniques such as Block Adaptive Quantization (BAQ)~\cite{lancashireBlockAdaptiveQuantization2001} and Flexible Dynamic BAQ (FDBAQ)~\cite{attemaFlexibleDynamicBlock2010a}, which operate on raw radar data and offer reliable, computationally efficient performance.
However, their compression efficiency remains limited compared to modern data-driven approaches.
In the computer vision community, Learned Image Compression (LIC) has recently achieved significant improvements in rate-distortion performance by optimizing compression pipelines end-to-end using deep neural networks~\cite{balleEndtoendOptimizedImage2017}.
Architectures based on variational autoencoders and hierarchical priors, such as the scale hyperprior model~\cite{balleVariationalImageCompression2018}, have become standard baselines for neural compression.

However, applying such techniques to SAR imagery is challenging due to the presence of multiplicative speckle noise.
Neural compression models may therefore allocate part of their representation capacity to modeling speckle rather than scene content, reducing compression efficiency.
Recent work has addressed this issue by jointly performing despeckling and compression~\cite{amao-olivaJointCompressionDespeckling2025}. 

Despite promising rate-distortion performance, deploying such architectures in operational spaceborne systems remains challenging.
Neural compression models typically require significant computational resources, making their deployment onboard processing units difficult, due to strict power and hardware constraints.
Field-Programmable Gate Arrays (FPGAs) are widely used in spaceborne payloads due to their radiation tolerance, energy efficiency, and flexibility for implementing specialized accelerators~\cite{NASASmallSpacecraftTechnology2015}.
However, adapting neural compression architectures to FPGA-based systems requires careful consideration of hardware constraints and operator compatibility~\cite{leonardFPGAEnabledMachineLearning2026}.

In this work, we investigate the requirements for making learned SAR image compression practical for onboard systems.
We use a joint SAR Despeckling and Data Compression (DDC) framework~\cite{amao-olivaJointCompressionDespeckling2025} as a case study and adapt it to an FPGA-based MPSoC.
Because the method operates on focused Single-Look Complex (SLC) products rather than raw echoes, a fully onboard pipeline would additionally require onboard focusing; we therefore treat this work as exploratory groundwork rather than a deployment-ready system.
Since existing DDC architectures were developed without deployment constraints, we investigate whether a deployment-aware redesign can preserve, or even improve, compression performance while enabling efficient embedded execution.

To this end, we perform a systematic ablation study over four model topologies, to identify suitable architectural choices for embedded execution. % including factorized-prior and scale-hyperprior variants with and without residual blocks.
Our experimental platform is a ZCU102 board using AMD's Vitis AI overlay accelerator~\cite{AMDVitisAI}.
Such a flexible accelerator is useful for prototyping.
Further, its practical limits, i.e., enforced fixed-point arithmetic and a restricted operator set, define the hardware-aware design space we study.
We therefore introduce DDC modifications that satisfy these constraints and evaluate the resulting models in both floating-point and integer precision across CPU, GPU, and FPGA platforms.

Our analysis shows that hardware-aware DDC models preserve---and even improve---rate-distortion performance, while offering a substantial energy advantage over GPU and CPU baselines.
Lastly, our efficiency study and onboard projection expose the remaining gaps to real onboard deployment.
The main contributions of this work are summarized as follows:

\begin{itemize}
    \item[\textbf{(C1)}] First work deploying joint SAR despeckling and learned compression on an FPGA-based embedded platform.
    \item[\textbf{(C2)}] Hardware-aware redesign demonstrating that deployment-driven modifications improve rate-distortion performance rather than degrading it.
    \item[\textbf{(C3)}] Systematic study revealing the performance-complexity tradeoff of architectural components (hyperprior, residual blocks, activation functions).
    \item[\textbf{(C4)}] Cross-platform characterization identifying the latency-energy tradeoffs relevant for onboard SAR processing.
    \item[\textbf{(C5)}] A projection of full-tile onboard processing and a discussion of the limits that remain towards a deployable onboard system.
\end{itemize}
This paper does not aim at establishing a new state-of-the-art compression algorithm.
Rather, it investigates the deployability of a learned joint compression-despeckling framework and, therefore, focuses on architectural and hardware aspects.

We introduce the background in Chapter~\ref{sec:background} and review related work in Chapter~\ref{sec:related}.
Chapter~\ref{sec:method} describes the proposed method and experimental setup, while Chapter~\ref{sec:results} presents and discusses the results.
We conclude with Chapter~\ref{sec:conclusion}.

\section{Background}\label{sec:background}

\subsection{SAR Data Characteristics and Compression}\label{sec:sar_background}

% ----- SAR differences from natural imaging -----
Synthetic Aperture Radar (SAR) sensors are active, coherent microwave imaging systems.
They emit microwave pulses whose echoes on the observed scene are recorded as complex signals containing amplitude and phase information.
As a result, SAR images differ from natural images in several ways.
Unlike passive sensors that rely on sunlight, observations can be made during the night.
Moreover, SAR wavelengths penetrate clouds, allowing all-weather observations.
This coherent imaging technique gives rise to speckle, a salt-and-pepper visual effect caused by the interference of the scattered echoes with each other.
While speckle can contain information about surface properties or movements, it also reduces image readability and limits visual interpretation.

% ----- Speckle for compression -----
In particular, the signal-dependent and multiplicative nature of speckle noise complicates the use of conventional DL methods.
Indeed, neural networks waste representation capabilities on noise rather than scene components.
This is worse in the case of data compression, as this noise also clutters the representation space of the compressed image.

\subsection{Learned Image Compression}\label{sec:lic_background}

% ----- End-to-end image compression using Neural Networks -----
Learned Image Compression (LIC) uses neural networks to perform \textit{lossy} data compression.
At the intersection of deep learning and information theory, LIC has been shown to outperform traditional handcrafted compression algorithms, such as JPEG or JPEG2000~\cite{balleEndtoendOptimizedImage2017}.

Conventional LIC architectures resemble autoencoders~\cite{hintonReducingDimensionalityData2006} and are trained end-to-end to jointly minimize the rate $\mathcal{R}$, the size of the compressed data, and the distortion $\mathcal{D}$, the difference between the reconstructed image and the original. % Like Variational Auto-Encoders (VAE)
In the loss $\mathcal{L}$, the tradeoff is controlled using a Lagrangian multiplier $\lambda$:

\begin{equation}
    \label{eq:RD-loss_simple}
    \mathcal{L} = \mathcal{R} + \lambda \cdot \mathcal{D}
\end{equation}

\begin{figure}[!t]
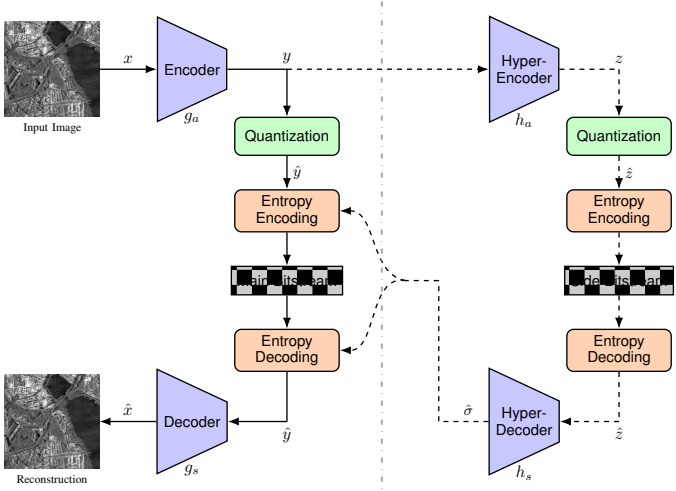

    \centering
    \includestandalone[width=\columnwidth]{figures/tikz/scale_hyperprior}
    \caption{Architecture diagrams of the Factorized Prior (\texttt{FP}) model, left of the dashed line only, and of the Scale Hyperprior (\texttt{SH}) model, the whole figure.}\label{fig:scale_hyperprior_architecture}
\end{figure}

A conventional LIC model, see Fig.~\ref{fig:scale_hyperprior_architecture}, uses an \textit{encoder} $g_a$ to map the image to a compressible latent space, which is quantized and losslessly encoded into a bitstream.
A \textit{decoder} $g_s$ can then invert this mapping to reconstruct the image.
The encoder $g_a$ and decoder $g_s$ serve as parametric alternatives to the \textit{analysis} and \textit{synthesis} transforms of classical codecs, e.g., the wavelet transform in JPEG2000~\cite{rabbaniBookReviewJPEG20002002}.
Training these autoencoder architectures requires no labels~\cite{balleVariationalImageCompression2018, minnenJointAutoregressiveHierarchical2018}.

% ----- Detailed training loss -----
The Rate-Distortion (RD) optimization, shown in Equation~(\ref{eq:RD-loss_simple}), expands as the expected bitstream length (the rate) and the reconstruction distortion:

\begin{equation}
    \label{eq:FP_rate}
    \displaystyle
    \mathcal{R}_{\mathrm{FP}} ~ = ~
    \mathbb{E}_{x \sim p_x}\left[-\log_2 p_{\hat{y}}(\lfloor g_a(x) \rceil )\right]
\end{equation}

\begin{equation}
    \label{eq:Distortion_detailed}
    \displaystyle
    \mathcal{D} ~ = ~
    \mathbb{E}_{x \sim p_x}\left[d\big(x, g_s(\lfloor g_a(x) \rceil)\big)\right]
\end{equation}

where $x$ is the input image and $p_x$ its associated, unknown distribution.
$\lfloor \cdot \rceil$ symbolizes the quantization operation, $y$ corresponds to the latents, and $p_{\hat{y}}$ is the discrete entropy model, with $\hat{y} = \lfloor y \rceil$.
The distortion $\mathcal{D}$ is computed using a distortion metric $d(\cdot)$, such as the Mean Squared Error (MSE) or Structural Similarity Index Measure (SSIM).
% The encoder $g_a$ transforms $x$, that follows an unknown distribution $p_x$, into its latent representation $y$.
% These latents are then quantized $\hat{y} = \lfloor y \rceil$ so that they can be modeled using a discrete entropy model $p_{\hat{y}}$.

% ----- Factorized prior model -----
Entropy modeling requires a prior probability model, for example, a fully factorized distribution that assumes no statistical dependencies in the latent distribution.
A variational autoencoder architecture using a fully factorized prior model was introduced by Ballé et al.~\cite{balleEndtoendOptimizedImage2017}.
We name this model Factorized Prior (\texttt{FP}), its architecture is depicted on the left half of Fig.~\ref{fig:scale_hyperprior_architecture}.

% ----- Hyperprior model -----
To enhance compression performance, one can introduce a new set of variables $z$ to capture the dependencies of the latent representations.
This side information can then be used to estimate $\hat{\sigma}$, the distribution of the standard deviations of each latent variable, which improves our entropy prior model.
In practice, this method requires two additional transformations implemented as neural networks: $h_a$, the \textit{hyper-encoder}, and $h_s$, the \textit{hyper-decoder}.
In this split-paradigm, the side information $z$ must also be quantized and compressed, demanding a second entropy model $p_{\hat{z}}$\footnote{As there are no prior beliefs about this distribution, a fully factorized density model is used.}.
With this additional information to encode, the rate $\mathcal{R}$ of Equation~(\ref{eq:FP_rate}) becomes:

\begin{equation}
    \label{eq:SH_rate}
    \mathcal{R}_{\mathrm{SH}} = \mathcal{R}_{\mathrm{FP}} + \mathbb{E}_{x \sim p_x}\left[-\log_2 p_{\hat{z}}(\lfloor h_a(y) \rceil )\right]
\end{equation}

The architecture using this new modeling of the latent distribution is referred to as Scale Hyperprior (\texttt{SH})~\cite{balleVariationalImageCompression2018} and is illustrated in Fig.~\ref{fig:scale_hyperprior_architecture}.
% \footnote{We implement the FP and SH models using the CompressAI framework~\cite{begaintCompressAIPyTorchLibrary2020} in which the models are respectively named \texttt{bmshj2018-factorized} and \texttt{bmshj2018-hyperprior}.}

\subsection{FPGA Acceleration for Onboard AI}\label{sec:fpga_background}
% % ----- Spaceborne constraints -----
% Spaceborne platforms face strict Size, Weight, Power, and Cost (SWaP-C) constraints, e.g., 
% At the same time, the space environment exposes electronics to Single-Event Effects (SEEs) caused by ionizing radiation~\cite{langeMachineLearningSpace2024}.

% ----- Accelerating NNs: platforms alternatives -----
Several hardware platforms are available when deploying neural networks onboard: CPUs, GPUs, ASICs, and specialized accelerators such as Vision Processing Units (VPUs).
CPUs have restricted parallel processing capabilities, limiting their efficiency on large computational workloads like neural networks.
In contrast, GPUs offer high peak throughput, but their power consumption often exceeds $300$\,W~\cite{collangePowerConsumptionGPUs2009}, making them impractical as SmallSat payloads typically operate within power budgets of $20$--$95$\,W~\cite{kothariFinalFrontierDeep2020}.
Some GPUs, like the NVIDIA Jetson Nano~\cite{NVIDIAJetsonNano}, are specially designed to run physically close to the sensor, off the power grid.
Such edge GPUs show reduced power consumption, but have not yet been designed with space-grade radiation reliability~\cite{leonardFPGAEnabledMachineLearning2026}, as ionizing radiation requires protecting electronics to avoid Single-Event Effects (SEEs)~\cite{langeMachineLearningSpace2024}.
Similarly, Intel's Myriad VPU~\cite{IntelMovidiusMyriad} has been shown to be unsuitable for long-duration Low Earth Orbit (LEO) missions due to radiation susceptibility~\cite{rapuanoFPGAbasedHardwareAccelerator2021a}.
Finally, ASICs can offer ultra-low power consumption and be radiation-hardened, but lack reconfigurability and are too long and expensive to design for most missions~\cite{amaraFPGAVsASIC2006}.

% ----- FPGAs in spaceborne payloads -----
Field-Programmable Gate Arrays (FPGAs) are reconfigurable integrated circuits that have become standard hardware in spaceborne payloads~\cite{siegleMitigationRadiationEffects2015, georgeOnboardProcessingHybrid2018, NASASmallSpacecraftTechnology2015}.
Their adoption in space systems is driven by several complementary properties.
FPGA vendors offer radiation-hardened product lines, and the reconfigurable fabric further enables software mitigation techniques~\cite{siegleMitigationRadiationEffects2015}.
Their reconfigurability also enables post-launch firmware updates and in-orbit reprogramming~\cite{leonardFPGAEnabledMachineLearning2026}.
Therefore, compared to other hardware alternatives, FPGAs remain the preferred choice in power- and space-constrained environments when radiation tolerance is required.
Moreover, with the growing adoption of AI acceleration on FPGAs, higher-level end-to-end toolchains have significantly reduced the barrier to deploying neural networks on FPGA hardware, such as FINN~\cite{umurogluFINNFrameworkFast2017} or Vitis AI~\cite{AMDVitisAI}.

\section{Related Work}\label{sec:related}

\subsection{SAR Data Compression}

% ----- RAW SAR compression: BAD, FDBAP -----
SAR data compression relies on traditional codecs like Block Adaptive Quantization (BAQ)~\cite{lancashireBlockAdaptiveQuantization2001} or its acquisition-dependent extension Flexible Dynamic BAQ (FDBAQ)~\cite{attemaFlexibleDynamicBlock2010a}, the current operational standard for the Copernicus Sentinel-1 mission.
FDBAQ adaptively allocates quantization bits based on local Signal-to-Noise Ratio (SNR) estimates and achieves variable bitrates while maintaining radiometric quality across heterogeneous scenes.
FDBAQ extends traditional BAQ by dynamically selecting from multiple quantizers depending on the backscatter intensity, resulting in improved performance over earlier entropy-constrained methods, particularly for bright targets, such as urban areas.

% ----- JPEG2000, and AI-BAQ -----
More recently, JPEG2000~\cite{rabbaniBookReviewJPEG20002002} has been evaluated for SAR raw data compression~\cite{asiyabiUseJPEG2000SAR2024}, demonstrating the superior performance of JPEG2000 compared to BAQ.
While these schemes are efficient, they remain handcrafted codecs using fixed transforms, and therefore have little adaptability to images.
AI-BAQ~\cite{gollinAIBAQDeepLearning2025} enhances the BAQ scheme in an acquisition-dependent scheme by using a lightweight Convolutional Neural Network (CNN) to predict bitrate maps for the raw data.
Such a data-driven strategy enhances the performance of BAQ without requiring onboard focusing or a priori scene information.

% ----- Further along the focusing chain -----
Further along the processing chain, LIC methods have been applied to focused SAR products.
Asiyabi et al.~\cite{asiyabiComplexValuedAutoencoderMultiPolarization2023} propose a complex-valued autoencoder architecture for compressing SLC data, leveraging dual-polarization channels as side information to improve reconstruction quality.
% \CL{If we submit the paper after EUSAR26 we should improve that section with the published methods there. At least MAYA\_DC but maybe other schemes presented there.}

\subsection{SAR Despeckling}\label{sec:related_sar}

All these methods consider compressing and reconstructing SAR images as a full signal, amplitude and phase.
However, allowing the loss of the phase information grants more flexibility during the representation learning. 
In particular, denoising can facilitate tasks like image compression~\cite{chengOptimizingImageCompression2022}, saving representation capabilities and representation space.
Removing speckle in SAR images is a long-tackled task, but has seen a recent breakthrough with a learning-based approach called MERLIN~\cite{dalsassoIfMagicSelfsupervised2022}.

One of the core problems of despeckling is the absence of ground-truth: no perfect SAR reflectivity image of a scene exists, making self-supervised approaches, like MERLIN, particularly attractive.
MERLIN exploits that SAR SLC real ($Re$) and imaginary ($Im$) parts are independent and identically distributed realizations of the same random process, allowing for a Noise2Noise approach~\cite{lehtinenNoise2NoiseLearningImage2018}.
Using the MERLIN training method, a model takes as input $Re^2$ (or $Im^2$) and estimates $Im^2$ (or $Re^2$).
It is trained to minimize Equation~(\ref{eq:MERLIN_loss}), where $\widehat{Re^2}$ is the prediction of the model and $k$ the index over the $N$ pixels of the images.

\begin{equation}
    \label{eq:MERLIN_loss}
    \mathcal{L}_{\mathrm{MERLIN}}(\widehat{Re^2}, Im^2) = \frac{1}{N}\sum_k \frac{1}{2} \log(\widehat{Re^2_k}) + \frac{Im^2_k}{\widehat{Re^2_k}}   
\end{equation}

Recent work from Amao-Oliva et al.~\cite{amao-olivaJointCompressionDespeckling2025} combines MERLIN despeckling with LIC.
We introduce the method further in Section~\ref{sec:method}.
However, while end-to-end despeckling and compression of SAR data enables fast quick-looking for disaster monitoring or various military applications, neural networks represent a significant computational load for a mission and must be optimized to suit onboard constraints.

\subsection{Neural Network Acceleration on FPGA}\label{sec:related_fpga}

% ----- Two broad approaches to NN acceleration on FPGAs -----
Designing an FPGA to accelerate neural networks can be achieved following two broad strategies. %~\cite{abdelouahabAcceleratingCNNInference2018, shawahnaFPGABasedAcceleratorsDeep2019}.
The first consists in writing a \textit{Specific} accelerator directly in Hardware Description Language (HDL) or High-Level Synthesis (HLS), tailoring the datapath precisely to a single network architecture.
This gives complete control over resource allocation, memory layout, and numerical precision, and has been the most common approach in the literature~\cite{leonardFPGAEnabledMachineLearning2026}.
% A specific accelerator can exploit mixed-precision quantization at the individual layer level, store weights entirely on-chip to eliminate memory-bandwidth bottlenecks, and implement multi-level parallelism---Instruction Level, Data Level, and Task Level---in a deeply pipelined dataflow fashion~\cite{umurogluFINNFrameworkFast2017, leonardFPGAEnabledMachineLearning2026}.
% For example, Yang et al.~\cite{yangAlgorithmHardwareCodesign2022} co-design architecture pruning and FPGA scheduling to achieve sub-millisecond SAR ship detection inference on a high-end Zynq UltraScale+, with on-chip storage of the compressed model weights.
For example, Sun et al.~\cite{sunFPGACodecSystem2024} co-optimize algorithm and FPGA design, constraining channel parallelism to improve the utilization of DSPs and therefore throughput.
However, HDL- or HLS-based design methods are low-level approaches with long, costly development cycles and require solid hardware expertise.
In addition, the resulting design cannot be reused for a different model without full reimplementation.
Nevertheless, hardware performance is significant, as Sun et al.~\cite{sunFPGACodecSystem2024} report an acceleration of $1.5\times$ the throughput of Chen et al.~\cite{chenExploringEfficientHardware2025}, which used the second strategy.

% ----- Flexible approach: Vitis AI DPU -----
The second design strategy relies on a \textit{Flexible} overlay accelerator, most prominently using AMD's Vitis AI~\cite{AMDVitisAI}.
Rather than building a circuit per model, a pre-synthesized Deep learning Processing Unit (DPU) is flashed to the FPGA once and can then execute any supported network by loading new weights from off-chip memory.
Such approach dramatically lowers the development barrier and has seen growing adoption for Remote Sensing applications~\cite{leonardFPGAEnabledMachineLearning2026}.
For example, Vitis AI has been used to accelerate a variety of CNNs used for cloud coverage detection~\cite{upadhyayDesignImplementationCNNbased2024}, land-cover segmentation~\cite{sabogalMethodologyEvaluatingAnalyzing2021a}, railway fault detection~\cite{yuImprovedLightweightDeep2024}, or multitask UAV monitoring~\cite{huangEdgeTrustworthyAI2024}.
As a drawback, using a fixed, generic deployment approach comes with its own constraints, such as an imposed 8-bit integer (\textbf{int8}) datatype and unsupported operators.
% For example, custom activation functions or entropy coding operations cannot be executed on the DPU and must be offloaded to the embedded ARM CPU present on the board.
% This introduces a DPU/CPU handoff latency that can dominate total inference time, as we observe in our results (Section~\ref{sec:results_compute}).

% ----- Placeholder: LIC acceleration on FPGA -----
% Possible search terms: "learned image compression FPGA", "neural image compression hardware", "autoencoder FPGA inference"
To our knowledge, the acceleration of LIC models on FPGAs has received limited attention to date.
Only few papers cover the deployment of LIC methods on FPGA, for video coding~\cite{jiaFPXNICFPGAAccelerated4K2022, chenExploringEfficientHardware2025}, or focus on a fixed network architecture acceleration~\cite{sunLearnedImageCodec2025, sunFPGACodecSystem2024}.
Notably, Mazouz et al.~\cite{mazouzLightweightEmbeddedFPGA2025} approach is the most related to our application.
However, none of these approaches apply to Remote Sensing or SAR data.

\section{Method}\label{sec:method}

In this section, we describe how we achieve our FPGA deployment and present the different models and environments used to evaluate
 the method.
To deploy the SAR Despeckling and Data Compression (DDC) framework on the ZCU102 Evaluation Kit we start from the original architecture presented by Amao-Oliva et al.~\cite{amao-olivaJointCompressionDespeckling2025}.

\subsection{A SAR Despeckling and Data Compression Framework}\label{sec:method_SAR_DDC}

% ----- Presentation SAR_DDC: introduce loss differences to MERLIN ----
SAR DDC is a self-supervised despeckling approach used to learn a SAR image representation that can then be used to perform image compression.
The model builds upon the MERLIN training strategy and incorporates LIC principles and objectives to perform speckle-free image compression and reconstruction.
In particular, the loss is adapted to keep the LIC rate term unchanged, see Equation~(\ref{eq:FP_rate}), but replaces the distortion term by the MERLIN objective of Equation~(\ref{eq:MERLIN_loss}).
For the \texttt{FP} variant \cite{balleEndtoendOptimizedImage2017}, this leads to:

\begin{equation}
    \label{eq:SAR_DDC_FP_loss}
    \mathcal{L}_{\mathrm{DDC,FP}} = \mathcal{R}_{\mathrm{FP}} + \lambda \cdot \mathcal{L}_{\mathrm{MERLIN}}
\end{equation}

For the \texttt{SH} variant~\cite{balleVariationalImageCompression2018}, the side-information rate term of Equation~(\ref{eq:SH_rate}) is added, yielding:

\begin{equation}
    \label{eq:SAR_DDC_SH_loss}
    \mathcal{L}_{\mathrm{DDC,SH}} = \mathcal{R}_{\mathrm{SH}} + \lambda \cdot \mathcal{L}_{\mathrm{MERLIN}}
\end{equation}

% ----- Introduce architectural differences to LIC -----
Regarding the architectures of the networks, the model used in the original study~\cite{amao-olivaJointCompressionDespeckling2025} follows the same structure as the \texttt{SH} network (see Fig.~\ref{fig:scale_hyperprior_architecture}) with a few tweaks.
Namely, it incorporates several residual blocks made of two convolutional layers separated by a Generalized Divisive Normalization (GDN) activation layer~\cite{balleDensityModelingImages2016}. %, no dimension reduction is made in these blocks.
These residual blocks are added after each GDN layer in the main encoder $g_a$ and decoder $g_s$.
The hyperprior networks $h_a$ and $h_s$ are left untouched.

\begin{figure}[!t]
    \centering
    \subfloat[\small Main encoder $g_a$]{%
        % ============================================================
% LAYOUT PARAMETERS — all in TikZ default units (cm).
% ============================================================
\def\gaW{0.30}         % Uniform width for ALL cuboids
\def\gaDepth{0.75}    % Isometric extrusion depth
\def\gaGap{0.55}      % Horizontal gap between consecutive cuboids
\def\gaHmax{4.0}      % Height at input resolution (256×256)
\def\gaHstep{0.55}    % Fixed height decrease per strided convolution
% ============================================================
\begin{tikzpicture}[font=\sffamily\scriptsize, >=Latex]

% ------------------------------------------------------------
% STEP 1 — Heights and y-bottoms (all blocks centred on y=0).
% ------------------------------------------------------------
\pgfmathsetmacro{\Hi}  {\gaHmax - 1*\gaHstep}   % Act1, Res1, Conv2
\pgfmathsetmacro{\Hii} {\gaHmax - 2*\gaHstep}   % Act2, Res2, Conv3
\pgfmathsetmacro{\Hiii}{\gaHmax - 3*\gaHstep}   % Act3, Res3, Conv4

\pgfmathsetmacro{\gaYzero} {-\gaHmax/2}
\pgfmathsetmacro{\gaYone}  {-\Hi/2}
\pgfmathsetmacro{\gaYtwo}  {-\Hii/2}
\pgfmathsetmacro{\gaYthree}{-\Hiii/2}

% Depth offsets for side-face arrow anchors.
% All blocks are centred on y=0, so the side-face y-centre = cbdy/2.
\pgfmathsetmacro{\cbdx}{\gaDepth*\isoCos}
\pgfmathsetmacro{\cbdy}{\gaDepth*\isoSin}
\pgfmathsetmacro{\arrowY}{\cbdy/2}

% ------------------------------------------------------------
% STEP 2 — X positions (left edge) and right edges of every block.
% ------------------------------------------------------------
\pgfmathsetmacro{\xCi}  {0.0}
\pgfmathsetmacro{\xAi}  {\xCi   + \gaW + \gaGap}
\pgfmathsetmacro{\xRi}  {\xAi   + \gaW + \gaGap}
\pgfmathsetmacro{\xCii} {\xRi   + \gaW + \gaGap}
\pgfmathsetmacro{\xAii} {\xCii  + \gaW + \gaGap}
\pgfmathsetmacro{\xRii} {\xAii  + \gaW + \gaGap}
\pgfmathsetmacro{\xCiii}{\xRii  + \gaW + \gaGap}
\pgfmathsetmacro{\xAiii}{\xCiii + \gaW + \gaGap}
\pgfmathsetmacro{\xRiii}{\xAiii + \gaW + \gaGap}
\pgfmathsetmacro{\xCiv} {\xRiii + \gaW + \gaGap}

\pgfmathsetmacro{\xCiR}  {\xCi   + \gaW}
\pgfmathsetmacro{\xAiR}  {\xAi   + \gaW}
\pgfmathsetmacro{\xRiR}  {\xRi   + \gaW}
\pgfmathsetmacro{\xCiiR} {\xCii  + \gaW}
\pgfmathsetmacro{\xAiiR} {\xAii  + \gaW}
\pgfmathsetmacro{\xRiiR} {\xRii  + \gaW}
\pgfmathsetmacro{\xCiiiR}{\xCiii + \gaW}
\pgfmathsetmacro{\xAiiiR}{\xAiii + \gaW}
\pgfmathsetmacro{\xRiiiR}{\xRiii + \gaW}
\pgfmathsetmacro{\xCivR} {\xCiv  + \gaW}

% ------------------------------------------------------------
% STEP 3 — Draw cuboids (no on-block labels; legend handles text).
% ------------------------------------------------------------
\drawcuboid{\xCi}  {\gaYzero} {\gaW}{\gaHmax}{\gaDepth}{conv_stride_orange}
\drawcuboidgdn{\xAi}  {\gaYone}  {\gaW}{\Hi}    {\gaDepth}
\drawcuboid{\xRi}  {\gaYone}  {\gaW}{\Hi}    {\gaDepth}{res_blue}
\drawcuboid{\xCii} {\gaYone}  {\gaW}{\Hi}    {\gaDepth}{conv_stride_orange}
\drawcuboidgdn{\xAii} {\gaYtwo}  {\gaW}{\Hii}   {\gaDepth}
\drawcuboid{\xRii} {\gaYtwo}  {\gaW}{\Hii}   {\gaDepth}{res_blue}
\drawcuboid{\xCiii}{\gaYtwo}  {\gaW}{\Hii}   {\gaDepth}{conv_stride_orange}
\drawcuboidgdn{\xAiii}{\gaYthree}{\gaW}{\Hiii}  {\gaDepth}
\drawcuboid{\xRiii}{\gaYthree}{\gaW}{\Hiii}  {\gaDepth}{res_blue}
\drawcuboid{\xCiv} {\gaYthree}{\gaW}{\Hiii}  {\gaDepth}{conv_stride_orange}

% ------------------------------------------------------------
% STEP 4 — Input / output arrows with labels above and dims below.
% ------------------------------------------------------------
\pgfmathsetmacro{\arrowLen}{0.8}
\pgfmathsetmacro{\xMidIn} {\xCi - \arrowLen/2}
\pgfmathsetmacro{\xMidOut}{\xCivR + \cbdx/2 + \arrowLen/2 + 0.20}

% x input arrow — tip at left front edge of Conv1
\draw[->, thick, black!65]
    ({\xCi - \arrowLen}, \arrowY) -- (\xCi, \arrowY);
\node[above=0.10cm, black!75] at (\xMidIn, \arrowY) {\Large $x$};
% \node[below=0.10cm, black!75] at (\xMidIn, \arrowY) {\tiny $256^2{\times}1$};

% y output arrow — tail at centre of Conv4's right face
\draw[->, thick, black!65]
    ({\xCivR + \cbdx/2}, \arrowY) -- ({\xCivR + \cbdx/2 + \arrowLen}, \arrowY);
\node[above=0.10cm, black!75] at (\xMidOut, \arrowY) {\Large $y$};
% \node[below=0.10cm, black!75] at (\xMidOut, \arrowY) {\tiny $16^2{\times}N$};

% % ------------------------------------------------------------
% % STEP 6 — Figure title.
% % ------------------------------------------------------------
% \node[below=0.6cm, align=center] at ({(\xCi + \xCivR)/2}, {-\gaHmax/2})
%     {\small Main analysis transform $g_a$
%      — downsampling $\times 16$,\quad $N=128$};

\end{tikzpicture}%
        \label{fig:ga_encoder}%
    }\\ %[6pt]
    \subfloat[\small Hyper-encoder $h_a$]{%
        % ============================================================
% LAYOUT PARAMETERS
% ============================================================
\def\haW{0.30}        % Uniform width for ALL cuboids
\def\haDepth{0.75}    % Isometric extrusion depth
\def\haGap{0.55}      % Horizontal gap between consecutive cuboids
\def\haHmax{3.0}      % Height at h_a input (= latent y from g_a, 16×16)
\def\haHstep{0.55}    % Height decrease per strided convolution
% ============================================================
\begin{tikzpicture}[font=\sffamily\scriptsize, >=Latex]

% ------------------------------------------------------------
% STEP 1 — Heights and y-bottoms (all blocks centred on y=0).
%   Conv1 (2N→N, k5s2) runs at full height.
%   After Conv1: Act1, Conv2 drop to Hi.
%   After Conv2: Act2, Conv3 drop to Hii.
% ------------------------------------------------------------
\pgfmathsetmacro{\haHi} {\haHmax - 1*\haHstep}   % Act1, Conv2
\pgfmathsetmacro{\haHii}{\haHmax - 2*\haHstep}   % Act2, Conv3

\pgfmathsetmacro{\haYzero}{-\haHmax/2}
\pgfmathsetmacro{\haYone} {-\haHi/2}
\pgfmathsetmacro{\haYtwo} {-\haHii/2}

% Side-face anchor offsets (same formula as ga_encoder)
\pgfmathsetmacro{\cbdx}{\haDepth*\isoCos}
\pgfmathsetmacro{\cbdy}{\haDepth*\isoSin}
\pgfmathsetmacro{\arrowY}{\cbdy/2}

% ------------------------------------------------------------
% STEP 2 — X positions (left edge) and right edges.
%   Order: Conv1 – Act1 – Conv2 – Act2 – Conv3
% ------------------------------------------------------------
\pgfmathsetmacro{\xCi}  {0.0}
\pgfmathsetmacro{\xAi}  {\xCi  + \haW + \haGap}
\pgfmathsetmacro{\xCii} {\xAi  + \haW + \haGap}
\pgfmathsetmacro{\xAii} {\xCii + \haW + \haGap}
\pgfmathsetmacro{\xCiii}{\xAii + \haW + \haGap}

\pgfmathsetmacro{\xCiR}  {\xCi   + \haW}
\pgfmathsetmacro{\xCiiiR}{\xCiii + \haW}

% ------------------------------------------------------------
% STEP 3 — Draw cuboids (no on-block labels; legend handles text).
% ------------------------------------------------------------
\drawcuboid{\xCi}  {\haYzero}{\haW}{\haHmax}{\haDepth}{conv_stride_orange}
\drawcuboid{\xAi}  {\haYone} {\haW}{\haHi}  {\haDepth}{act_violet}
\drawcuboid{\xCii} {\haYone} {\haW}{\haHi}  {\haDepth}{conv_stride_orange}
\drawcuboid{\xAii} {\haYtwo} {\haW}{\haHii} {\haDepth}{act_violet}
\drawcuboid{\xCiii}{\haYtwo} {\haW}{\haHii} {\haDepth}{conv_stride_orange}

% ------------------------------------------------------------
% STEP 4 — Input / output arrows with labels above and dims below.
% ------------------------------------------------------------
\pgfmathsetmacro{\arrowLen}{0.8}
\pgfmathsetmacro{\xMidIn} {\xCi - \arrowLen/2}
\pgfmathsetmacro{\xMidOut}{\xCiiiR + \cbdx/2 + \arrowLen/2 + 0.20}

% y input arrow — tip at left front edge of Conv1
\draw[->, thick, black!65]
    ({\xCi - \arrowLen}, \arrowY) -- (\xCi, \arrowY);
\node[above=0.10cm, black!75] at (\xMidIn, \arrowY) {\Large $y$};
% \node[below=0.10cm, black!75] at (\xMidIn, \arrowY) {\tiny $16^2{\times}2N$};

% z output arrow — tail at centre of Conv3's right face
\draw[->, thick, black!65]
    ({\xCiiiR + \cbdx/2}, \arrowY) -- ({\xCiiiR + \cbdx/2 + \arrowLen}, \arrowY);
\node[above=0.10cm, black!75] at (\xMidOut, \arrowY) {\Large $z$};
% \node[below=0.10cm, black!75] at (\xMidOut, \arrowY) {\tiny $4^2{\times}N$};

% % ------------------------------------------------------------
% % STEP 5 — Figure title.
% % ------------------------------------------------------------
% \node[below=0.6cm, align=center] at ({(\xCi + \xCiiiR)/2}, {-\haHmax/2})
%     {\Large Hyper analysis transform $h_a$
%      --- downsampling $\times 4$,\quad $N=128$};

\end{tikzpicture}%
        \label{fig:ha_encoder}%
    }\hfill%
    \subfloat[\small Residual block]{%
        % ============================================================
% LAYOUT PARAMETERS
% ============================================================
\def\rbW{0.30}        % Uniform block width
\def\rbDepth{0.75}    % Isometric extrusion depth
\def\rbGap{0.55}      % Gap between cuboids
\def\rbH{3.0}         % All blocks the same height (no stride)
% ============================================================
\begin{tikzpicture}[font=\sffamily\scriptsize, >=Latex]

% y-bottom (blocks centred on y=0)
\pgfmathsetmacro{\rbY}{-\rbH/2}

% Side-face offsets
\pgfmathsetmacro{\cbdx}{\rbDepth*\isoCos}
\pgfmathsetmacro{\cbdy}{\rbDepth*\isoSin}
\pgfmathsetmacro{\arrowY}{\cbdy/2}

% X positions
\pgfmathsetmacro{\rbxCone}  {0.0}
\pgfmathsetmacro{\rbxAone}  {\rbxCone + \rbW + \rbGap}
\pgfmathsetmacro{\rbxCtwo}  {\rbxAone + \rbW + \rbGap}

\pgfmathsetmacro{\rbxConeR} {\rbxCone + \rbW}
\pgfmathsetmacro{\rbxCtwoR} {\rbxCtwo + \rbW}

% Merge point (circled +): just right of Conv2's right side-face
\pgfmathsetmacro{\arrowLen}{0.8}
\pgfmathsetmacro{\xMerge}{\rbxCtwoR + \cbdx + \arrowLen}
\def\circR{0.18}   % radius of the + circle

% Skip connection top: above the isometric top face
\pgfmathsetmacro{\skipTop}{\rbH/2 + \cbdy + 0.40}

% Arrow lengths and midpoints for labels
\pgfmathsetmacro{\xMidIn} {\rbxCone - \arrowLen/2}
\pgfmathsetmacro{\xMidOut}{\xMerge + \circR + \arrowLen/2}

% ------------------------------------------------------------
% Draw cuboids
% ------------------------------------------------------------
\drawcuboid{\rbxCone}{\rbY}{\rbW}{\rbH}{\rbDepth}{conv_orange}
\drawcuboid{\rbxAone}{\rbY}{\rbW}{\rbH}{\rbDepth}{act_violet}
\drawcuboid{\rbxCtwo}{\rbY}{\rbW}{\rbH}{\rbDepth}{conv_orange}

% ------------------------------------------------------------
% Skip connection: 3-segment path (up, across, down) into ⊕
% ------------------------------------------------------------
\draw[thick, black!55]
    (\rbxCone - \arrowLen/2, \arrowY)
    -- (\rbxCone - \arrowLen/2, \skipTop)
    -- (\xMerge, \skipTop)
    -- (\xMerge, {\arrowY + \circR});

% Circled + node at merge point
\node[circle, draw=black!55, thick, inner sep=2pt, fill=white]
    (plus) at (\xMerge, \arrowY) {\small $+$};

% ------------------------------------------------------------
% Input arrow (→ into left front face of Conv1)
% ------------------------------------------------------------
\draw[->, thick, black!65]
    ({\rbxCone - \arrowLen}, \arrowY) -- (\rbxCone, \arrowY);

% Main-path arrow: Conv2 right side-face → ⊕
\draw[->, thick, black!65]
    ({\rbxCtwoR + \cbdx/2}, \arrowY) -- (plus.west);

% Output arrow: ⊕ → rightward
\draw[->, thick, black!65]
    (plus.east) -- ({\xMerge + \circR + \arrowLen}, \arrowY);

% % ------------------------------------------------------------
% % Figure title
% % ------------------------------------------------------------
% \node[below=0.6cm, align=center] at ({(\rbxCone + \xMerge)/2}, \rbY)
%     {\small Residual block};

\end{tikzpicture}%
        \label{fig:resblock_detail}%
    }\\ %[6pt]
    \subfloat[\small Block legend]{%
        % ============================================================
% LAYOUT PARAMETERS
% ============================================================
\def\lgW{0.20}       % Block width  (thin slab)
\def\lgDepth{0.45}   % Isometric depth
\def\lgH{0.75}       % Height = Depth → square front face
\def\lgHgap{1.30}    % Horizontal pitch between swatch left-edges
\def\lgLabelY{0.12}  % Drop below block bottom for the text label
\def\lgLabelX{0.2}  % Drop below block bottom for the text label
% ============================================================
\begin{tikzpicture}[font=\sffamily\scriptsize, >=Latex]

\pgfmathsetmacro{\cbdx}{\lgDepth*\isoCos}
\pgfmathsetmacro{\cbdy}{\lgDepth*\isoSin}

% X left-edges (five swatches, evenly pitched)
\pgfmathsetmacro{\xConvS}{0*\lgHgap}
\pgfmathsetmacro{\xConv} {1*\lgHgap}
\pgfmathsetmacro{\xGDN}  {2*\lgHgap}
\pgfmathsetmacro{\xReLU} {3*\lgHgap}
\pgfmathsetmacro{\xRes}  {4*\lgHgap}

% Label anchor = centre of front face
\pgfmathsetmacro{\lbConvS}{\xConvS + \lgW/2}
\pgfmathsetmacro{\lbConv} {\xConv  + \lgW/2}
\pgfmathsetmacro{\lbGDN}  {\xGDN   + \lgW/2}
\pgfmathsetmacro{\lbReLU} {\xReLU  + \lgW/2}
\pgfmathsetmacro{\lbRes}  {\xRes   + \lgW/2}

% All swatches share the same y-baseline
\pgfmathsetmacro{\yBase}{0}

% --- Strided Conv (darker orange) ---
\drawcuboid{\xConvS}{\yBase}{\lgW}{\lgH}{\lgDepth}{conv_stride_orange}
\node[below=\lgLabelY cm, align=center] at (\lbConvS + \lgLabelX, \yBase)
    {Conv\\$k=5$\\$s=2$};

% --- Plain Conv (warm orange) ---
\drawcuboid{\xConv}{\yBase}{\lgW}{\lgH}{\lgDepth}{conv_orange}
\node[below=\lgLabelY cm, align=center] at (\lbConv + \lgLabelX, \yBase)
    {Conv\\$k=5$\\$s=1$};

% --- GDN or ReLU activation (striped violet) ---
\drawcuboidgdn{\xGDN}{\yBase}{\lgW}{\lgH}{\lgDepth}
\node[below=\lgLabelY cm, align=center] at (\lbGDN + \lgLabelX, \yBase)
    {GDN\\or\\ReLU};

% --- Plain ReLU activation (solid violet) ---
\drawcuboid{\xReLU}{\yBase}{\lgW}{\lgH}{\lgDepth}{act_violet}
\node[below=\lgLabelY cm, align=center] at (\lbReLU + \lgLabelX, \yBase)
    {ReLU};

% --- Residual block (steel blue) ---
\drawcuboid{\xRes}{\yBase}{\lgW}{\lgH}{\lgDepth}{res_blue}
\node[below=\lgLabelY cm, align=center] at (\lbRes + \lgLabelX, \yBase)
    {Residual\\block (c)};

\end{tikzpicture}%
        \label{fig:arch_legend}%
    }
    \caption{DDC encoder-side architecture.\ (a)~Main analysis transform $g_a$, downsampling $\times16$.\ (b)~Hyper analysis transform $h_a$, downsampling $\times4$.\ (c)~Structure of each residual block.\ (d)~Legend of the different blocks: $k$ represents the kernel size (square) and $s$ the stride.}\label{fig:architecture_overview}
\end{figure}
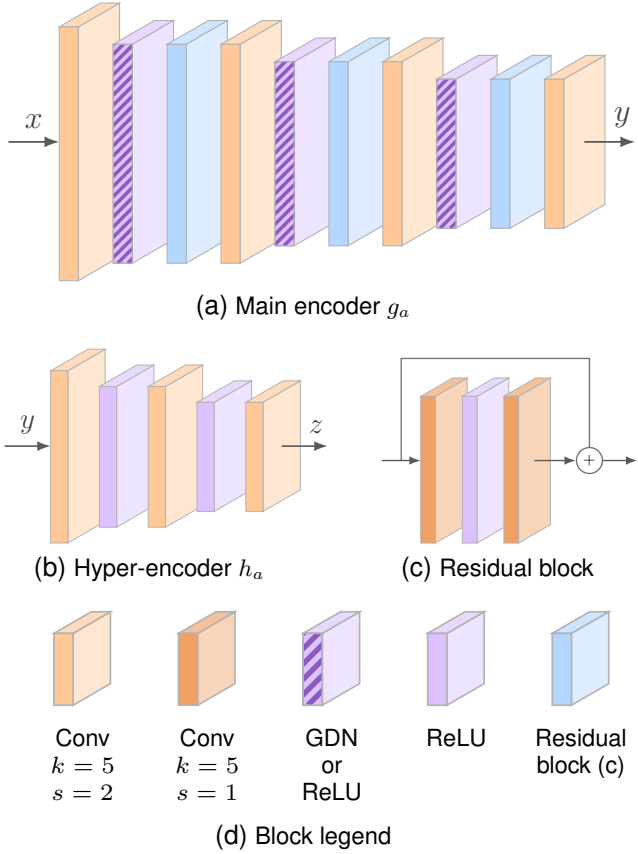

The exact architectures of the main encoder, the hyper-encoder, and the residual blocks are depicted in Fig.~\ref{fig:architecture_overview}.
We omit the decoder and hyper-decoder architectures for clarity, as they mirror the encoder and hyper-encoder.
Specifically, $g_s$ and $h_s$ replace convolutional downsampling layers with transposed convolutions that progressively upsample the feature maps back to the original image dimensions.

\begin{figure*}
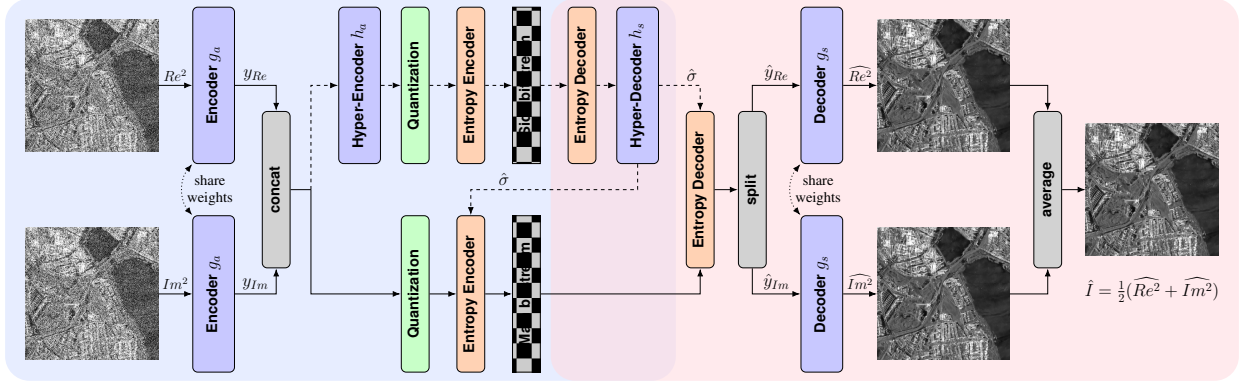

    \centering
    \includestandalone[width=0.9\linewidth]{figures/tikz/SAR_DDC_inference_dataflow}
    \caption{DDC inference dataflow. The left of the figure, (blue background) shows the \textit{compression} of the image, while the right of the figure (pink background) represents the \textit{decompression}. In the case of a Hyperprior architecture, denoted with dashed arrows, the Entropy Decoder and the Hyper-Decoder (purple background) must be executed for both scenarios.}\label{fig:SAR_DDC_inference_dataflow}
\end{figure*}

% ----- Explain the 2 forward passes for Real/Imaginary parts -----
One additional modification to the original \texttt{SH} network made by DDC, is the doubling of the latent space size (and consequently of the number of input and output channels of $h_a$ and $h_s$).
Indeed, while MERLIN uses only one of the real ($Re$) or imaginary ($Im$) representation of the image during training (see Section~\ref{sec:related_sar}), during inference DDC uses a single forward pass to compress $Re$ and $Im$, as seen in Fig.~\ref{fig:SAR_DDC_inference_dataflow}.
The main encoder $g_a$ and decoder $g_s$ are executed twice, and the latent representation $y_{Re}$ and $y_{Im}$ are concatenated to leverage the data redundancies and allow a unique forward pass through $h_a$ and $h_s$.

\subsection{Different Model Architectures}\label{sec:model_architectures}

As the original architecture from Amao-Oliva et al.~\cite{amao-olivaJointCompressionDespeckling2025} uses residual blocks with the Scale Hyperprior (\texttt{SH}) model, we refer to it as \texttt{ResSH}.
In addition, we evaluate three other model configurations: \texttt{SH}, \texttt{FP}, and \texttt{ResFP}, using the Factorized Prior (\texttt{FP}) model, with and without residual blocks (see Fig.~\ref{fig:ga_encoder}).
In Section~\ref{sec:results_arch}, we systematically evaluate these four existing topologies under the DDC framework in terms of compression quality and computational complexity.
% In Section~\ref{sec:results_arch}, we perform a comparison of these four model topologies in terms of compression quality and computational complexity.

Table~\ref{tab:network_stats} describes the computational complexity of the four architectures.
\textit{\#Params} corresponds to the total number of parameters, \textit{Mem.} to the networks' memory footprint in 32-bit floating-point (\textbf{float32}) precision, and \#OPs to the workload, in numbers of operations, necessary to process a $256 \times 256$ patch. % (the DPU uses the 8-bit integer (\textbf{int8}) datatype, thus with a footprint $\approx \frac{1}{4}$)
This workload is reported for the full inference scenario depicted on Fig.~\ref{fig:SAR_DDC_inference_dataflow} and only for the compression scenario, i.e., blue background.

\begin{table}[t]
    \centering
    \setlength{\tabcolsep}{4pt}
    \caption{Computational complexity of the four architectures.}\label{tab:network_stats}
        \begin{tabular}{l S[table-format=3.3] S[table-format=3.3] S[table-format=3.3] S[table-format=3.3]}
        \toprule
        {Arch.} & {\#Params [M]} & {Mem.\,[MB]} & {\#OPs\,full\,[G]} & {\#OPs\,comp.\,[G]}\\
        \midrule
        \texttt{FP}    & 2.06  & 8.22  & 14.80  & 9.02  \\
        \texttt{ResFP} & 7.02  & 28.08 & 155.77 & 79.51 \\
        \texttt{SH}    & 10.14 & 40.57 & 15.70  & 9.57  \\
        \texttt{ResSH} & 15.11 & 60.42 & 156.68 & 80.06 \\
        \bottomrule
        \end{tabular}
\end{table}

Table~\ref{tab:subnetwork_stats} gives a more fine-grained analysis of the workload and reports the statistics of each of the subnetworks used in these architectures.
Here, Max DPU FPS corresponds to the isolated single-core ceiling, i.e., the highest possible throughput for a subnetwork without any overhead; it is not equivalent to the end-to-end throughput, but dictates  the highest possible FPS of the entire model (as these networks must execute sequentially).
% \CL{I'm undecided if I keep this Max DPU FPS column, it might be too hardware for TGRS. On the other hand, I find it interestig because it defines the compute boundary for such a method, i.e., because these networks must execute sequentially, the highest possible FPS of the entire model (even if using perfect memory transfer, pipelining, etc.) is $28 FPS \approx 35 ms$ for Res architectures. I let you decide Paolo.} % Maybe also need to define it in caption, it's not obvious what it means.

\begin{table}[t]
    \centering
    \setlength{\tabcolsep}{4pt}
    \caption{Computational complexity of subnetworks.}\label{tab:subnetwork_stats}
    \begin{tabular}{l S[table-format=3.3] S[table-format=3.3] S[table-format=3.3] S[table-format=3.3]}
        \toprule
        {Subnet} & {\#Params [M]} & {Mem.\,[MB]} & {\#OPs\,[G]} & {Max DPU FPS} \\
        \midrule
        $g_a$ + Res & 3.73 & 14.93 & 39.75 & 27.8  \\
        $g_s$ + Res & 3.29 & 13.16 & 38.13 & 28.5  \\
        \midrule
        $g_a$       & 1.25 &  5.00 &  4.51 & 219.4 \\
        $g_s$       & 0.81 &  3.23 &  2.89 & 266.1 \\
        \midrule
        $h_a$       & 4.93 & 19.71 &  0.28 & 1225.7 \\
        $h_s$       & 3.16 & 12.63 &  0.18 & 1374.7 \\
        \bottomrule
    \end{tabular}
\end{table}

\subsection{Data and Training Setup}\label{sec:method_setup}

Our dataset is made from five TerraSAR-X StripMap (SM) SSC (Single Look Slant Range Complex) tiles\footnote{The images were obtained via ESA TerraSAR-X archive at \url{https://earth.esa.int/eogateway/catalog/terrasar-x-esa-archive}}.
The images are split into $\sim 35{,}000$ $256 \times 256$ patches to create training ($63\%$), evaluation ($21\%$) and testing ($16\%$) datasets following a spatial split.

Following the MERLIN self-supervised training strategy, training requires no labels.
However, we use the U-Net architecture~\cite{ronnebergerUNetConvolutionalNetworks2015} presented in the original MERLIN study by Dalsasso et al.~\cite{dalsassoIfMagicSelfsupervised2022} to generate despeckled images of the scenes for evaluation and testing.
As this reference is itself an estimate, the resulting metrics should be read comparatively across models rather than as absolute fidelity.
% \CL{Maybe we should discuss and motivate why we use MERLIN and not temporally-averaged coregistered acquisitions.
% Also, I have no comparison for the compression performance, like MERLIN + JPEG in the original paper. I think we agreed that we don't do any solid comparison because its not the job of this paper and instead rely on what was shown in your previous work. But maybe we should explicitly state that somewhere?}

The training configurations used throughout this work are summarized in Table~\ref{tab:training_setup}.
All models are implemented using CompressAI~\cite{begaintCompressAIPyTorchLibrary2020}, trained with the Adam optimizer~\cite{kingmaAdamMethodStochastic2017}, with a batch size of $12$, and use gradient-norm clipping set to $1.0$.

\begin{table}[t]
    \centering
    \caption{Training configurations used in this work.}\label{tab:training_setup}
    % Deatiled caption: Training configurations used in this work. Framework refers to the training approach, Arch.\ to the model architecture, Res. Blocks if the architecture contains residual blocks (see Fig.~\ref{fig:resblock_detail}), Loss to the equation minimized during training, $\mathrm{lr}$ to the learning rate, Epochs to the number of epochs each model was trained for, and Runs how many models were trained using different seeds.
    \footnotesize
    \setlength{\tabcolsep}{3pt}
    \renewcommand{\arraystretch}{1.15}
    \begin{tabular}{c c c c c c c}
        \toprule
        Framework & Arch. & Res. Blocks & Loss & $\mathrm{lr}$\tablefootnote{Following MERLIN~\cite{dalsassoIfMagicSelfsupervised2022} procedure, learning rates are reduced by a factor 10 at epoch $4$ and $20$. For each model, the initial optimal learning rate was selected after a grid hyperparameter search $lr \in [10^{-5}; 10^{-3}]$.} & \#Epochs & \#Runs \\
        \midrule
        MERLIN & U-Net & Yes & (\ref{eq:MERLIN_loss}) & $10^{-4}$ & 30 & 1 \\
        \midrule
        DDC & \texttt{FP} & No & (\ref{eq:SAR_DDC_FP_loss}) & $5 \cdot 10^{-4}$ & 10 & 6 \\
        DDC & \texttt{ResFP} & Yes & (\ref{eq:SAR_DDC_FP_loss}) & $5 \cdot 10^{-4}$ & 10 & 6 \\
        DDC & \texttt{SH} & No & (\ref{eq:SAR_DDC_SH_loss}) & $5 \cdot 10^{-4}$ & 10 & 6 \\
        DDC & \texttt{ResSH} & Yes & (\ref{eq:SAR_DDC_SH_loss}) & $10^{-4}$ & 10 & 6 \\
        \bottomrule
    \end{tabular}
\end{table}

Each architecture is trained for ten different RD-tradeoffs, i.e., $\lambda$ values from $1$ to $1{,}000$, and for six different runs (seeds).
RD-curves show results averaged across runs and standard deviations are displayed as vertical error bands (distortion) and horizontal error bars (rate).
Because the complete test set was too large to fit on the ZCU102, FPGA results are evaluated on a representative subset of the complete test set, namely $500$ images out of $5{,}724$.
The complete code, methodological details, and additional experiments are available at \url{https://github.com/CedricLeon/SAR_DDC_FPGA}.

\subsection{Deployment Strategy}

% ----- Vitis AI DPU -----
After training and evaluation, we deploy suitable models to the FPGA system~(\textbf{C1}).
In this work, we use the AMD Vitis AI toolchain~\cite{AMDVitisAI} to implement SAR DDC on an AMD Zynq UltraScale+™ MPSoC ZCU102 Evaluation Kit---containing sizable FPGA logic and an ARM Cortex-A53 CPU.

Vitis AI implements a \textit{Flexible} design approach: a pre-built Deep Learning Processing Unit (DPU)\footnote{AMD has renamed this accelerator to Neural Processing Unit (NPU) in their latest releases. However, in Vitis AI 3.0, used in this study, the accelerator was still called DPU. We use the DPU B4096 configuration.} is flashed onto the FPGA and can execute different neural network workloads without requiring FPGA reconfiguration.
However, the DPU is designed for \textbf{int8} datatype operations and requires quantizing the model weights, which we accomplish using the Post-Training Quantization (PTQ) method.
% Vitis AI deployment process involves model weights quantization, which is realized using the Post-Training Quantization (PTQ) method, perturbing the model parameters away from their convergence point.
% The process results in different metrics when running inference in \textbf{float32} on CPU or GPU or in \textbf{int8} on the DPU.

In addition, the DPU only supports a restricted set of operators; layers that fall outside this set must be executed on the embedded ARM CPU instead.
This DPU-to-CPU constraint, together with the \textbf{int8} quantization requirement, directly motivates the hardware-aware architectural modifications we describe next.

\subsection{DDC Hardware-Aware Modifications}\label{sec:method_modifications}

The modifications outlined in this section constitute the necessary hardware-aware redesign~(\textbf{C2}) required for models to comply with the limited set of operations supported by the DPU.
The original \texttt{ResSH-GDN} model has three architectural features that violate these rules:
\begin{enumerate}
    \item The entropy encoding and decoding operations are not supported by AMD's DPU.
    \item The GDN~\cite{balleDensityModelingImages2016} activation function is unknown to Vitis AI.
    \item Transposed convolutions kernels used in the main and hyper-decoder $g_s$ and $h_s$ use non-zero output padding, which is unofficially unsupported by the DPU.
\end{enumerate}

We tackle these problems by reshaping transposed convolution kernels from $(5\times5, 2, 1)$ to $(4\times4, 1, 0)$, defined as $(k \times k, ip, op)$, where $k$ refers to the size of the square kernel, $ip$ to the input padding and $op$ to the output padding.
In addition, we replace GDN activations, present in $g_a$ and $g_s$, by Rectified Linear Unit (ReLU) activations.

% The absence of entropy-coding support in standard DPU-based toolchains is a key barrier, since entropy encoding and decoding are integral to LIC pipelines but fall outside the operator sets of existing overlay accelerators.
% This gap directly motivates the DPU/CPU subgraph design we adopt in this work.

Finally, we segment the inference forward pass into DPU and CPU subgraphs: the neural transforms $g_a$, $h_a$, $h_s$, and $g_s$ run on the Vitis AI DPU, while the entropy encoding and decoding run on the ARM CPU.
Such a breakdown of the computational graph is bound to introduce transfer overhead, but also gives the flexibility to implement various execution pipelines.
For example, as $g_a$ and $g_s$ must be executed twice to process the real and imaginary parts, we utilize the several cores of the Vitis AI DPU B4096 to parallelize them, i.e., $g_a(Re^2) \parallel g_a(Im^2)$ (same applies to $g_s$).
% \CL{Maybe it's worth replacing the description of the dataflow CPU/DPU and parallelization by a small figure... But we have quite some figures already.}

\section{Results}\label{sec:results}

% In this section we evaluate DDC's behavior on the ZCU102, from the impact of the hardware-aware modifications to its computational efficiency using diverse model architectures and onboard potential.
In this section, we progressively analyze the deployment of the DDC framework on the ZCU102: from algorithmic modifications to architectural choices, and finally hardware efficiency.

\subsection{Impact of Hardware-Aware Modifications}\label{sec:results_ablation}

\begin{figure}
    \centering
    \includegraphics[trim={0.25cm 0.3cm 0.25cm 0.2cm},clip,width=\linewidth]{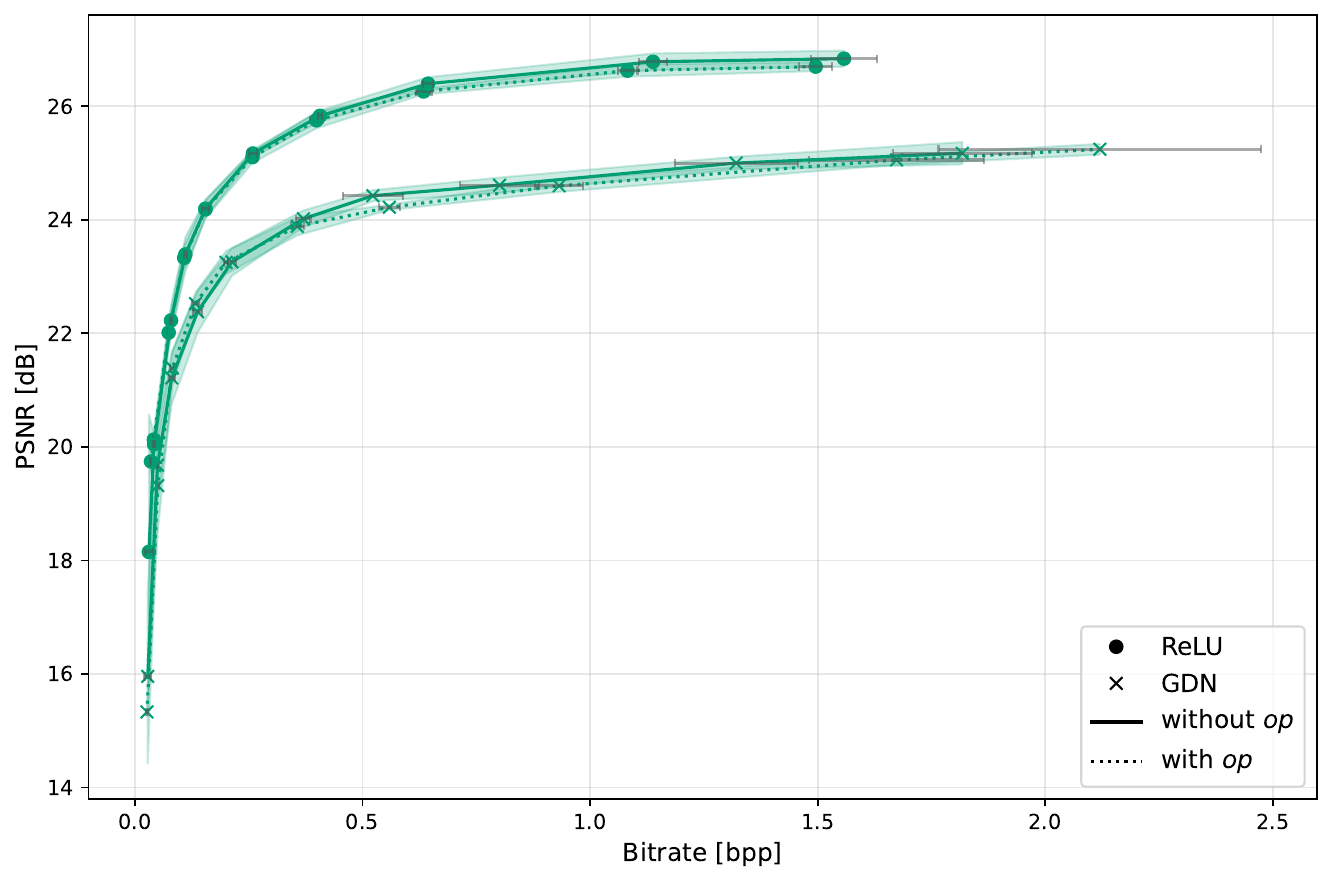}
    \caption{FPGA-aware modifications of the \texttt{ResSH} architecture, in \textbf{float32} precision. Each RD-curve has varying activation functions (ReLU / GDN, cf.\ Fig.~\ref{fig:architecture_overview}) and transposed-convolution output-padding (with / without).}\label{fig:ablation_RD}
    % Detailed caption: Each RD-curve has varying activation functions (ReLU / GDN, cf.\ Fig.~\ref{fig:architecture_overview}) and transposed-convolution output-padding (with / without).
\end{figure}

Due to DPU constraints, the DDC algorithm, and its original \texttt{ResSH} architecture, cannot run on the ZCU102. 
In this first experiment, we analyze the impact of the necessary modifications to deploy the model on the FPGA-based MPSoC~(\textbf{C2}).
Fig.~\ref{fig:ablation_RD} shows the ablation study over the original SAR DDC architecture \texttt{ResSH} in \textbf{float32} precision.
% The figure illustrates the rate-distortion (RD) tradeoff: minimize the bpp on the X-axis while maximizing the Peak Signal-to-Noise Ratio (PSNR) on the Y-axis.

Notably, replacing GDN by ReLU improves the RD tradeoff by $1.67\pm0.10$\,dB and $0.26\pm0.07$\,bpp at the highest evaluated rate ($\lambda=1{,}000$). % SSIM improvements: +0.0442±0.0024
% Such result is unexpected as GDN~\cite{balleDensityModelingImages2016} has been specifically developed to decorrelate and Gaussianize feature maps, therefore, improving image modelling and minimizing redundancies across latent variable for a better entropy coding.
Such a result is unexpected, as GDN was designed to Gaussianize feature activations and improve the efficiency of density modeling, proving especially effective for LIC~\cite{balleDensityModelingImages2016}.
In contrast, ReLU simply introduces non-linearity and is not tailored for LIC.
We hypothesize that SAR data characteristics, in particular speckle and the Rayleigh distribution followed by intensity images are significantly different from natural images and hinder GDN improvements. %\CL{(Is that still true after the normalization we make? The normalization is log + Min-Max.)} 

Reducing the size of the inverse convolution kernels to $(4\times4)$ also positively impacts the results, though only marginally.
Overall, both modifications come at no performance cost---and even surprisingly improve the \texttt{ResSH} baseline---while, most importantly, enabling us to deploy the DDC framework on the ZCU102 board.

\subsection{Architecture Exploration: Performance vs.\ Complexity}\label{sec:results_arch}

\begin{figure}
    \centering
    \includegraphics[trim={0.25cm 0.3cm 0.25cm 0.2cm},clip,width=\linewidth]{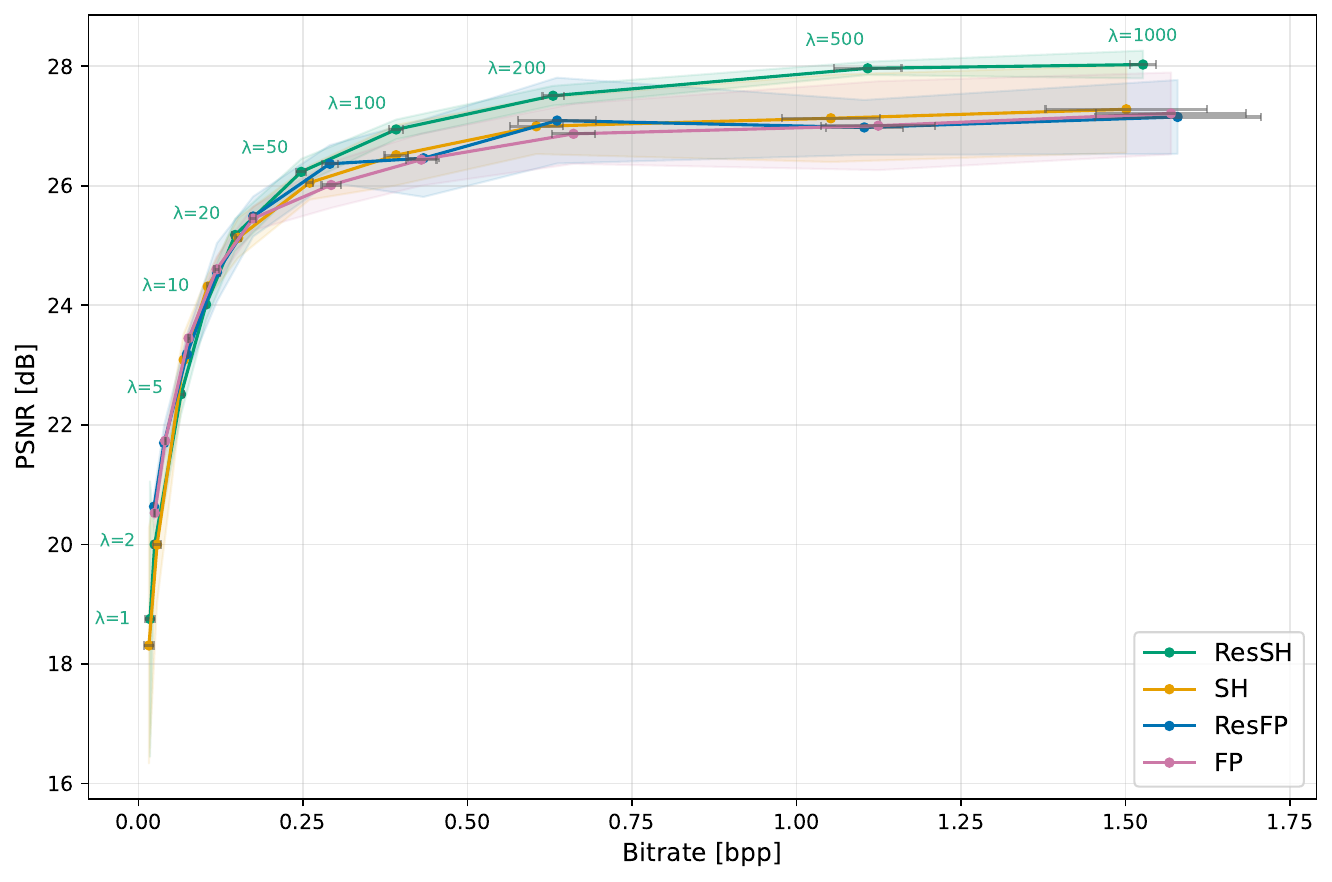}
    \caption{RD performance of four model topologies. % (\texttt{ResSH}, \texttt{SH}, \texttt{ResFP}, \texttt{FP}) of the DDC framework.
    All architectures are evaluated in \textbf{float32} precision and use the FPGA-ready modifications (no output padding in the decoders and ReLU activations).}\label{fig:rd_4arch}
\end{figure}

In this second experiment, we aim at identifying the onboard-suitability of the original \texttt{ResSH} architecture and characterize the cost-benefit of its specific architectural components~(\textbf{C3}).
To this end, we evaluate four model architecture alternatives: \texttt{ResSH}, \texttt{SH}, \texttt{ResFP}, and \texttt{FP}.
Fig.~\ref{fig:rd_4arch} shows their comparison at \textbf{float32} precision.

At first glance all four architectures seem to behave similarly, with a light domination of \texttt{ResSH}, the original DDC architecture, at higher bitrates.
However, such analysis does not take into account the differences in computational complexity highlighted in Table~\ref{tab:network_stats}.
Indeed, using an architecture with residual blocks has a significant cost overhead: \texttt{ResSH} requires $156.68$\,GOPs to compress a patch, while \texttt{SH} needs $15.70$\,GOPs.
As the improvement of \texttt{ResSH} over \texttt{SH} ($0.75\pm0.31$\,dB at $\lambda = 1{,}000$) is consistent but modest, we observe that the residual blocks' $10 \times$ compute overhead buys little quality in return.
Looking further, \texttt{SH}, \texttt{FP}, and \texttt{ResFP} perform similarly at higher rates, favoring \texttt{FP} as its workload and mem
ory footprint are the lowest, see Table~\ref{tab:network_stats}.
All architectures perform similarly at low rates until the knee of the RD-curve ($\lambda=20$).
Consequently, if a single model had to be chosen for onboard deployment, we recommend the lightweight \texttt{FP} architecture.

\subsection{The Cost of Quantization}\label{sec:results_crossprec}

\begin{figure*}
    \centering
    \includegraphics[trim={0.1cm 0.25cm 0.2cm 0.2cm},clip,width=\linewidth]{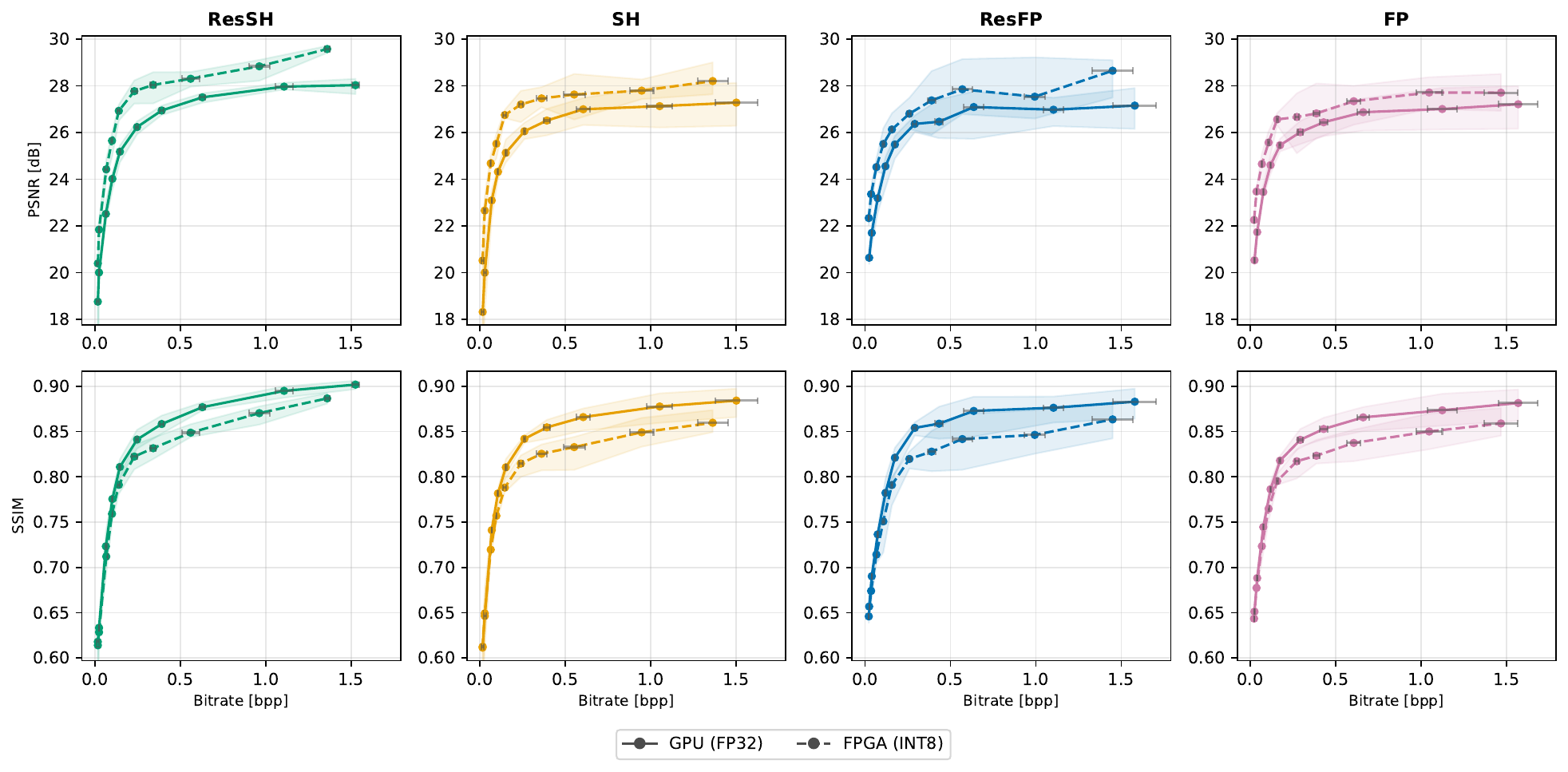}
    \caption{RD-curves of the four architectures across two different hardware platforms and precision GPU (\textbf{float32}) vs.\ FPGA (\textbf{int8}). The top row shows the PSNR and the bottom row the SSIM.}\label{fig:crossprec_rd}
\end{figure*}

As per DPU-requirements, models need to be adapted before deployment.
In particular, their parameters need to be quantized from \textbf{float32} to \textbf{int8} datatype, reducing numerical precision and saving computational costs and memory.
Fig.~\ref{fig:crossprec_rd} quantifies the resulting rate-distortion cost of such a deployment-driven modification~(\textbf{C2}), for each architecture and for two distortion metrics. 

We see that for all architectures the FPGA \textbf{int8} quantized model scores higher than the full-precision GPU baseline, by $+0.50$ (\texttt{FP}) to $+1.55$\,dB (\texttt{ResSH}) at the highest rate.
Such results are counter-intuitive as quantization (especially PTQ) inevitably shifts model parameters from their convergence point and causes accuracy degradation compared to full-precision implementations~\cite{gholamiSurveyQuantizationMethods2021}.
However, looking at a perceptual metric like the SSIM tells a different story with a quantization cost of $\approx 0.02$ points.

One possible interpretation is that \textbf{int8} reconstructions are systematically less bright than their \textbf{float32} counterparts, matching more closely the despeckled reference across the dark majority of the scenes while penalising the sparse high-scatterer areas.
Overall, this leads to desaturated images that lower pixel-wise errors, such as MSE (on which PSNR is based).
We also observe quantization effects as structure perturbations, but it is more nuanced.
While SSIM, that relies on local brightness and contrast statistics, barely drops, the Edge Preservation Degree (EPD), a purely gradient-based metric, drops by $0.25$ (\texttt{ResSH}) to $0.33$ (\texttt{FP}).
% We also observe quantization effects as a structure perturbation with the small SSIM drop, but specifically associated with edge displacement as the Edge Preservation Degree (EPD), a gradient-based metric, drops ranging from $0.25$ (\texttt{ResSH}) to $0.33$ (\texttt{FP}).
This consistent drop, larger in EPD than SSIM, seems to indicate that quantization preserves the energy around edges, but degrades their localization.

\begin{figure}
    \centering
    \includegraphics[trim={0.25cm 0.3cm 0.25cm 0.2cm},clip,width=\linewidth]{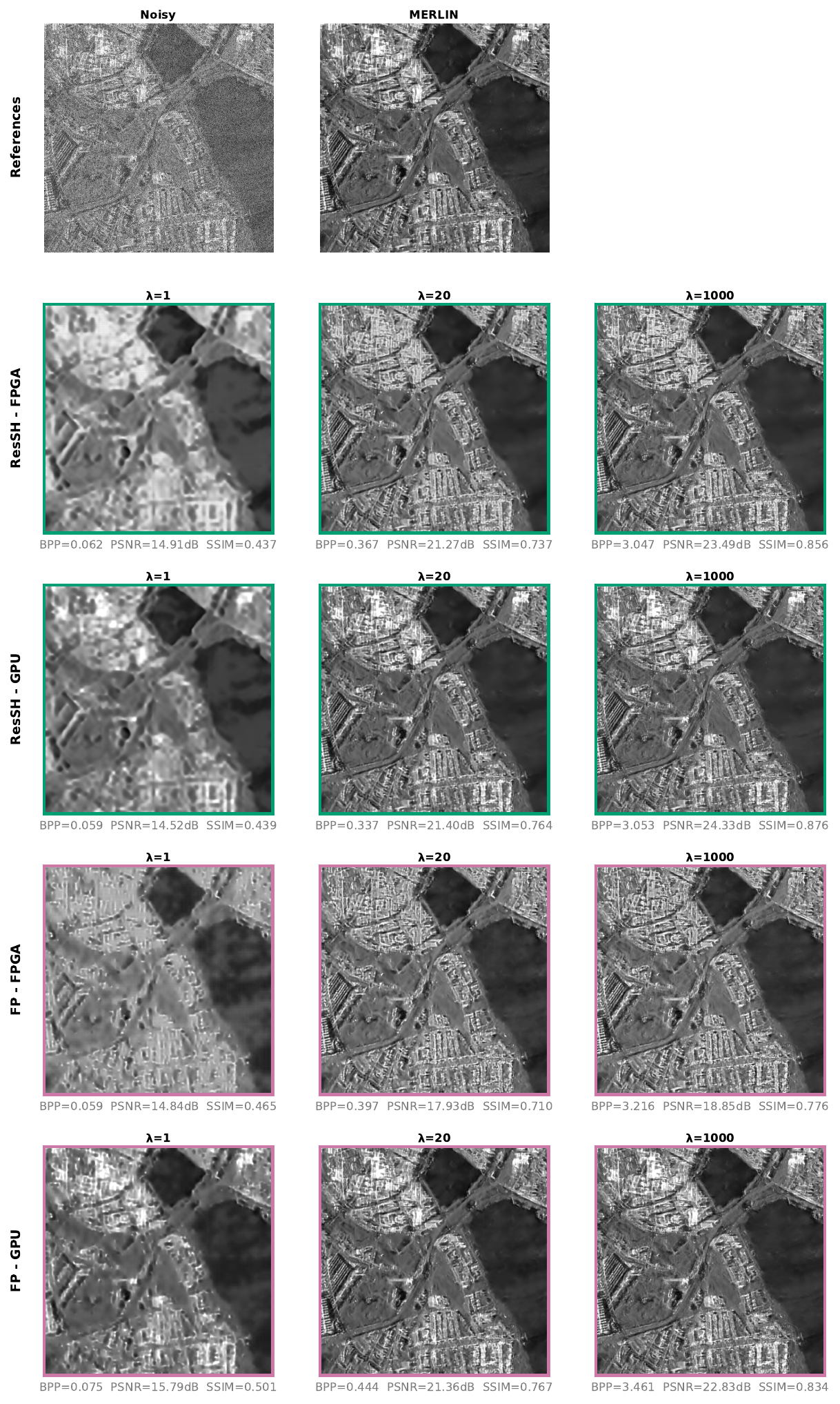}
    \caption{Comparison of reconstructions over Hamburg city-center ($1{,}024 \times 1{,}024$). First row, noisy SLC and MERLIN-despeckled reference. Subsequent rows show \texttt{ResSH} and \texttt{FP}, each on GPU (\textbf{float32}) and FPGA (\textbf{int8}), for three rate points ($\lambda = 1, 20, 1000$). Metrics (Bitrate [bpp], PSNR [dB], and SSIM) are annotated per tile.}\label{fig:qual_grid}
\end{figure}

Fig.~\ref{fig:qual_grid} visualizes these quantization-induced effects.
The first row displays the input and reference images, the original noisy SLC and the reflectivity image despeckled using MERLIN.
The other rows show DDC reconstructions at different rate points for two different architectures \texttt{ResSH} and \texttt{FP} and at two different precisions, the original \textbf{float32} model on GPU and the adapted \textbf{int8} model deployed on the FPGA.

We chose the three rate points ($\lambda = 1, 20, 1000$) to be about an order of magnitude apart in bitrate, spanning Compression Ratios (CR) of $\sim 1{,}650$, $215$, and $23$ on average on the test set (FPGA).
% Calculation: Original = FP32 x 2 channels -> reconstruction = bpp x 1 channel:
%    bpp=0.05 -> CR=1280
%    bpp=0.25 -> CR=256
%    bpp=2.5 -> CR=25.6
% Also, as the tile is 1024x1024 pixels, the bpp corresponds to the final image size in Megabits, e.g., Lambda=1 gives an image of 0.05 Mb or 6.4 KyloBytes.
Aside from the net increase in visual quality from very-low rates ($\lambda=1$) to the knee of the RD-curve ($\lambda=20$), it is interesting to see how much less saturated FPGA reconstructions are in high-scatterer areas compared to the GPU. % notice that FPGA models tend to produce less saturated reconstructions in high-scatterer areas compared to the GPU.
Once again, we attribute this behavior to the \textbf{int8} output-range cap, ultimately leading to smoother images.
This observation is also consistent with the lower FPGA bitrates and SSIM values between GPU and FPGA.
Note, however, that this image of Hamburg city-center tile is a high-scatterer scene, where the reduced \textbf{int8} range is most penalizing: \texttt{FP} scores a lower PSNR than its \textbf{float32} counterpart, contrary to the test-set average of Fig.~\ref{fig:crossprec_rd}, while the more quantization-robust \texttt{ResSH} still shows a small gain.

\subsection{Computational Efficiency}\label{sec:results_compute}

While \texttt{ResSH} is superior to \texttt{FP} in terms of RD performance, maximizing computational efficiency is also key for onboard processing~(\textbf{C4}).
Table~\ref{tab:network_stats} already shows that \texttt{ResSH} and \texttt{ResFP} architectures have a workload $10 \times$ higher than their counterparts without residual blocks.
In terms of memory footprint, \texttt{FP} is about $5\times$ smaller than \texttt{SH}, highlighting that the hyperprior path adds two large networks $h_a$ and $h_s$ but these process small inputs, resulting in minimal workload increase.

\begin{figure}
    \centering
    \includegraphics[trim={0.25cm 0.3cm 0.25cm 0.25cm},clip,width=\linewidth]{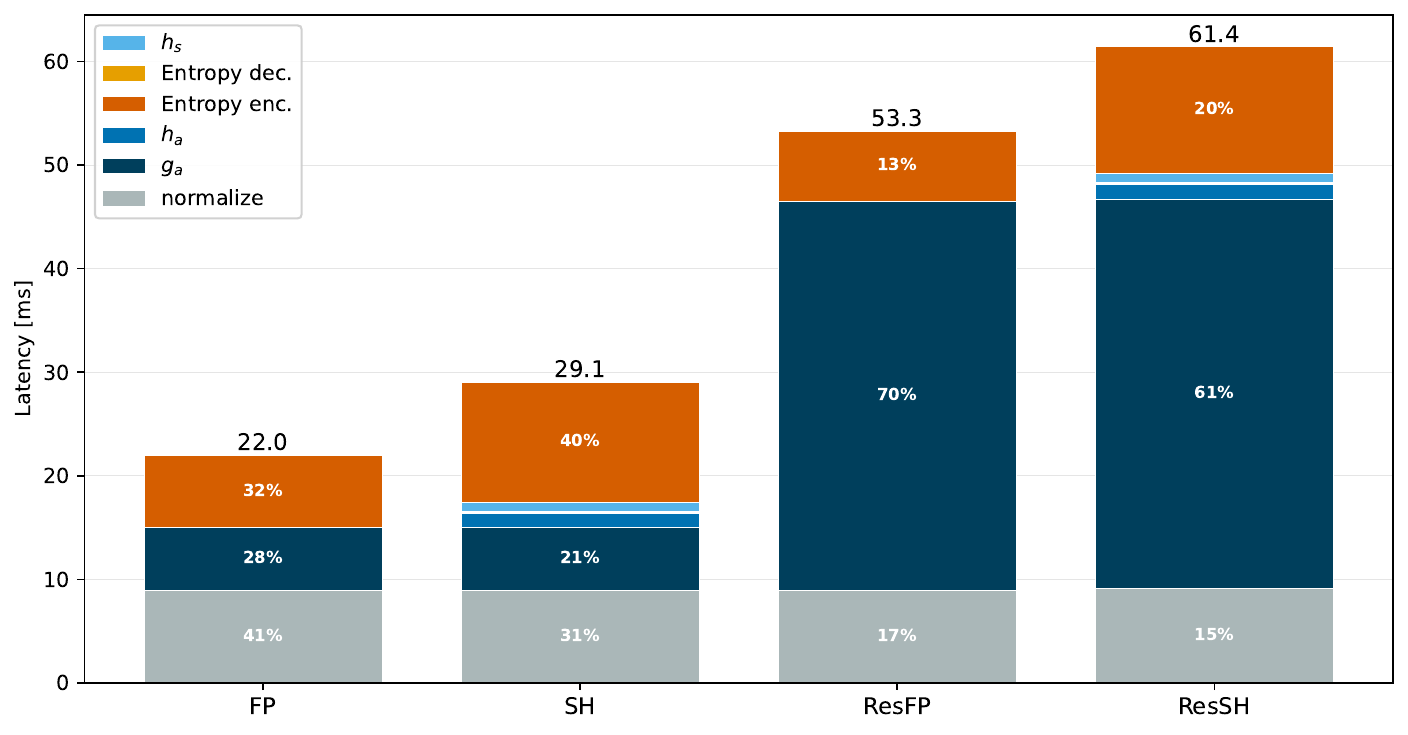}
    \caption{Breakdown of the latencies of each architecture to compress one patch on the FPGA.}\label{fig:latency_breakdown}
\end{figure}

However, such theoretical reasoning does not always translate into practice, as the implementation on the ZCU102 using the Vitis AI DPU includes overheads such as preprocessing, memory transfers, etc.
Therefore, Fig.~\ref{fig:latency_breakdown} analyzes the latencies of the different stages of each architecture and shows that \texttt{FP}, the computationally-cheapest model, has the lowest latency.
It is interesting to notice that the overhead from the log-normalization of the image (executed on un-vectorized ARM CPU) is constant across all archs ($\sim 9.02$\,ms) and reaches up to $41\%$ of the compute time for \texttt{FP}.

We can also see the costs of the different entropy encoding/decoding schemes.
E.g., \texttt{SH} and \texttt{ResSH} architectures perform two encodings, one in the hyperprior path with a negligible cost (given the minimal size of the hyper-latent $z$) and a second one in the main path.
This second encoding uses the predicted scales $\hat{\sigma}$ to perform the compression and has almost twice the cost of models using a simple factorized prior ($6.7$\,ms vs.\ $11.9$\,ms).
Finally, comparing our current parallelized implementation of $g_a(Re^2)$ and $g_a(Im^2)$ with a sequential approach execution shows a saving of $4.4$\,ms ($42.0\%$) for $g_a$, or $35.6$\,ms ($48.7\%$) for $g_a$ with residual blocks, demonstrating an excellent use of the DPU cores.

Table~\ref{tab:crossplatform_energy_latency} compares latency and energy per patch against two desktop platforms: an Intel Xeon w3--2423 CPU and an NVIDIA RTX A4000 GPU.
Predictably, the GPU achieves the lowest latency across all architectures, as the CPU and FPGA are dominated by the large computational cost of the encoder $g_a$.

\subsection{Energy Efficiency}\label{sec:results_energy}

Energy efficiency is particularly important for onboard operation, where power budgets and thermal dissipation are binding constraints~(\textbf{C4}).
Here, the embedded FPGA is significantly more frugal, consuming $11$--$17\times$ less energy per patch than the CPU and $2$--$6\times$ less than the GPU.
The ZCU102 MPSoC draws $9.8$--$12.7$\,W of active power against $108$--$126$\,W for the A4000\footnote{Measured power scopes differ slightly and should only be used as indications, see the associated repository for details.}, comfortably within the $20$--$95$\,W budget typical of SmallSat payloads~\cite{kothariFinalFrontierDeep2020}. % (FPGA MPSoC vs.\ GPU board+CPU vs.\ CPU package+DRAM)
This frugality is what makes the full-tile projection we present next compelling, despite the FPGA's modest per-patch latency.

\begin{table}[t]
    \centering
    \caption{Latency and energy consumption per patch across different hardware platforms for the compression scenario, $\times$ vs.\ CPU.}\label{tab:crossplatform_energy_latency}
    \setlength{\tabcolsep}{6pt}
    \begin{tabular}{c c S[table-format=6.1] l S[table-format=6.1] l}
        \toprule
        {Arch.} & {Platform} & \multicolumn{2}{c}{Latency\,[ms]} & \multicolumn{2}{c}{Energy\,[mJ]} \\
        \midrule
        \multirow{3}{*}{\texttt{FP}} & CPU & 26.7 & \textemdash & 2341.8 & \textemdash \\
         & GPU & 7.4 & $3.6\times$ & 866.5 & $2.7\times$ \\
         & FPGA & 22.0 & $1.2\times$ & 214.5 & $11\times$ \\
        \midrule
        \multirow{3}{*}{\texttt{SH}} & CPU & 40.4 & \textemdash & 3487.1 & \textemdash \\
         & GPU & 16.1 & $2.5\times$ & 1738.6 & $2.0\times$ \\
         & FPGA & 29.1 & $1.4\times$ & 284.6 & $12\times$ \\
        \midrule
        \multirow{3}{*}{\texttt{ResFP}} & CPU & 113.6 & \textemdash & 10951.9 & \textemdash \\
         & GPU & 9.3 & $12\times$ & 1175.6 & $9.3\times$ \\
         & FPGA & 53.3 & $2.1\times$ & 659.5 & $17\times$ \\
        \midrule
        \multirow{3}{*}{\texttt{ResSH}} & CPU & 126.3 & \textemdash & 12360.7 & \textemdash \\
         & GPU & 15.1 & $8.4\times$ & 1862.8 & $6.6\times$ \\
         & FPGA & 61.4 & $2.1\times$ & 782.5 & $16\times$ \\
        \bottomrule
    \end{tabular}
\end{table}

\subsection{End-to-End Onboard Projection}\label{sec:results_e2e}

To put these per-patch figures in an operational context, we extrapolate them to a complete TerraSAR-X SLC tile, gauging the potential and limits of an onboard deployment~(\textbf{C5}).
Our test tile of $21{,}036\times29{,}828$ resolution cells yields $11{,}000$ overlapping $256^2$ patches (16-px overlap to avoid edge artifacts, as performed for Fig.~\ref{fig:qual_grid}).
Compressing it would take $11.3$ minutes and $8.6$\,kJ with \texttt{ResSH}, or $4.0$ minutes and $2.36$\,kJ with \texttt{FP}.
The $2.51$\,GB tile (real and imaginary stored as \textbf{int16}) shrinks to $123$--$132$\,MB (CR $19.0$--$20.5$) at the highest rate, or $12.7$--$14.2$\,MB (CR $177$--$198$) near the RD knee, a one-to-two order-of-magnitude reduction in downlink volume.
While these are only first-order estimates that ignore potential inter-patch scheduling and memory constraints, they illustrate the throughput and downlink savings within reach of an onboard implementation.

\subsection{Limits and Future Work}\label{sec:limits_and_future_work}

While successfully deploying the SAR DDC application on FPGAs unlocks many onboard applications, we see several limits to considering the DDC framework onboard future missions. % and characterizing architectural tradeoffs 
First of all, it lies at a very specific spot in the traditional SAR processing pipeline: it requires fully focused SLC images and produces non-complex despeckled images.
Onboarding such a method would, therefore, require the presence of a focuser onboard.
In this regard, while real-time implementations exist~\cite{mandapatiRealTimeFloating2024}, it would still be a significant mission-design commitment to allocate the hardware and resources to focus, compress, and despeckle acquired data.
Second, while the \textit{Flexible} Vitis AI DPU~\cite{AMDVitisAI} allows us to easily experiment with various architectures, it is likely suboptimal, as it was not designed for LIC.
More specifically, we do not make use of all the FPGA logic resources available and CPU-DPU transfers result in non-negligible overhead processing.

Nonetheless, as this work is not carried out to achieve the best RD-tradeoff nor aims for real-time performance, we see great optimization potential for this family of methods.
% In particular, we expect significant perceptual loss recovery over using Quantization-Aware Training (QAT) instead of PTQ as seen in the literature~\cite{papatheofanousSoCFPGAAcceleration2022}.
We did not experiment with using Quantization-Aware Training (QAT) instead of PTQ, as it is non-trivial to implement for our partitioned CPU-DPU computational graph.
However, QAT represents a promising direction for future work as we can expect significant perceptual loss recovery~\cite{papatheofanousSoCFPGAAcceleration2022}.
We also see significant room for improvement in terms of hardware performance.
Specifically, a \textit{Specific} FPGA design could significantly improve hardware performance.
In particular, implementing the log-normalization of patches on the programmable logic of the ZCU102 should considerably lighten the computational burden of the ARM CPU.
Another promising direction would be to investigate a streaming pipeline in which patches are transferred from memory and processed continuously across multiple compute stages, allowing the performance assessment of full-tile processing.
Finally, beyond performance optimization, the speckle-free and compact latent learned by DDC enables a further direction: attaching task-specific decoders (for instance for detection or segmentation) directly to this latent would enable onboard exploitation without first decompressing the image.

\section{Conclusion}\label{sec:conclusion}

In this work, we presented the first deployment of a joint SAR Despeckling and Data Compression (DDC) framework on an embedded FPGA.
The implementation used the Vitis AI overlay accelerator and required hardware-aware adaptations of the original DDC architecture to comply with the DPU's 8-bit integer datatype and restricted operator set.
We replaced the GDN activations and output-padded transposed convolutions, and partitioned the computational graph into DPU and CPU subgraphs, mapping the neural transforms onto the DPU while offloading entropy coding to the ARM core.
Beyond satisfying the accelerator's constraints, this partition exposed a practical parallelization opportunity, as the two real and imaginary passes through the main encoder and decoder can be dispatched concurrently across the DPU cores.
Building on this deployment, and motivated by the absence of any such study in the original framework, we carried out a systematic ablation across four DDC topologies and complemented it with a cross-precision and cross-platform efficiency analysis spanning CPU, GPU, and FPGA.

Several findings emerged from this characterization.
Replacing GDN with ReLU improved the rate-distortion tradeoff rather than degrading it, suggesting that the simpler non-linearity suits the speckle statistics of SAR better than the density-modeling operator borrowed from natural-image compression.
The architecture ablation further revealed that the larger residual variants buy little reconstruction quality for an order of magnitude more computation, so that, if a single model had to be selected, the lightweight factorized-prior variant would be preferred for embedded deployment.
Across precisions, integer quantization traded numerical range for more homogeneous reconstructions, raising PSNR while lowering perceptual fidelity, a behavior consistent across all four model architectures. % and that should be minded when using pixel-wise metrics to assess image reconstruction.
Finally, despite its modest per-patch latency, the embedded FPGA proved by far the most energy-efficient of the three platforms. %, which is precisely the property that matters most under the power budgets of spaceborne payloads.
More broadly, our results suggest that deployment-aware neural network design should be considered an integral part of learned SAR compression rather than a post-processing engineering step.
% Together, these results map both the promise and the open gaps of accelerating learned SAR compression on reconfigurable hardware, and we hope they offer a grounded reference for the architectural, precision, and efficiency choices that future onboard designs will have to make.
We hope these findings offer a grounded reference for the architectural and hardware choices that future onboard designs will have to make.

\section*{Acknowledgments}
This work is supported by the Helmholtz Association under the joint research school ``Munich School for Data Science - MUDS''. % chktex 8.
This work benefited from the use of Claude Sonnet 4.6, which provided alternative formulations of some parts of the manuscript.

\bibliographystyle{IEEEtran}
\bibliography{references/TGRS}

% \section*{Appendices}

\end{document}